\documentclass[journal]{vgtc}                     % final (journal style)

\usepackage{amsmath}
\usepackage{enumitem}

\onlineid{0}

\vgtccategory{Research}

\vgtcpapertype{theory/model}

\title{Perceiving Slope and Acceleration: Evidence for Variable Tempo Sampling in Pitch-Based Sonification of Functions}

\author{%
  \authororcid{Danyang Fan}{0000-0002-3368-7887}
  \authororcid{Walker Smith}{0009-0000-5850-4728}
  \authororcid{Takako Fujioka}{0000-0003-2421-9106}
  \authororcid{Chris Chafe}{0000-0002-5381-0109}
  \authororcid{Sile O'Modhrain}{0000-0003-3804-5469}
  \authororcid{Diana Deutsch}{0000-0002-6060-1196}
  \authororcid{Sean Follmer}{0000-0001-5592-5949}
}

\abstract{%
  Sonification offers a non-visual way to understand data, with pitch-based encodings being the most common. Yet, how well people perceive slope and acceleration---key features of data trends---remains poorly understood. Drawing on people's natural abilities to perceive tempo, we introduce a novel sampling method for pitch-based sonification to enhance the perception of slope and acceleration in univariate functions. While traditional sonification methods often sample data at uniform x-spacing, yielding notes played at a fixed tempo with variable pitch intervals (Variable Pitch Interval), our approach samples at uniform y-spacing, producing notes with consistent pitch intervals but variable tempo (Variable Tempo). We conducted psychoacoustic experiments to understand slope and acceleration perception across three sampling methods: Variable Pitch Interval, Variable Tempo, and a Continuous (no sampling) baseline. In slope comparison tasks, Variable Tempo was more accurate than the other methods when modulated by the magnitude ratio between slopes. For acceleration perception, just-noticeable differences under Variable Tempo were over 13 times finer than with other methods. Participants also commonly reported higher confidence, lower mental effort, and a stronger preference for Variable Tempo compared to other methods. This work contributes models of slope and acceleration perception across pitch-based sonification techniques, introduces Variable Tempo as a novel and preferred sampling method, and provides promising initial evidence that leveraging timing can lead to more sensitive, accurate, and precise interpretation of derivative-based data features.
}


\graphicspath{{figs/}{figures/}{pictures/}{images/}{./}} % where to search for the images

\usepackage{tabu}                      % only used for the table example
\usepackage{booktabs}                  % only used for the table example
\usepackage{lipsum}                    % used to generate placeholder text
\usepackage{mwe}                       % used to generate placeholder figures

\usepackage{mathptmx}                  % use matching math font

\begin{document}

%%%%%%%%%%%%%%%%%%%%%%%%%%%%%%%%%%%%%%%%%%%%%%%%%%%%%%%%%%%%%%%%
%%%%%%%%%%%%%%%%%%%%%% START OF THE PAPER %%%%%%%%%%%%%%%%%%%%%%
%%%%%%%%%%%%%%%%%%%%%%%%%%%%%%%%%%%%%%%%%%%%%%%%%%%%%%%%%%%%%%%%

\firstsection{Introduction}

\maketitle

Data sonification---the process of representing data through sound---offers a compelling alternative to visual displays \cite{walker2011theory}, particularly for supporting real-time monitoring \cite{hildebrandt2016continuous, gaver1988everyday, mauney2004creating}, enhancing accessibility \cite{noel2022accessibility, sharif2022makes, potluri2022psst}, promoting pattern recognition \cite{Parvizi_2018}, and fostering public engagement and outreach \cite{zanella2022sonification, lindborg2023climate}. Among the auditory dimensions used in sonification, pitch is the most common \cite{dubus_systematic_2013, hermann2011sonification}. Previous studies have evaluated pitch-based techniques for tasks such as magnitude estimation \cite{wang2022seeing, walker2002magnitude}, feature identification (e.g., extrema) \cite{fan2022slide}, and trend detection \cite{harrar2007designing, nees2008data}. Beyond these tasks, slope and acceleration play a central role in interpreting change in functional relationships between variables, whether estimating growth, identifying inflection points, or recognizing nonlinear patterns. Yet, in contrast to the visual domain---where slope judgments have been extensively studied \cite{talbot2012empirical, parrott2014spatial, ciccione2021can, cleveland1987graphical}---relatively little research has examined how sonification supports perception of slope and acceleration.

% While natural language approaches like summaries and alternative text can offer valuable high-level context \cite{lundgard2021accessible}, direct perceptual mappings provide a complementary approach by granting transparent, low-effort access to underlying data patterns without requiring an interpretive agent \cite{fan2023accessibility, setlur2016eviza}.

For pitch-based sonification of continuous univariate functions (i.e., $y = f(x)$), common sampling approaches include mapping data to a continuous pitch trajectory, or sampling data at uniform time intervals to produce notes at a fixed tempo with varying pitches \cite{harrar2007designing, nees2008data, sharif2022makes, hermann2011sonification}. In the context of this study, we refer to these methods as Continuous and Variable Pitch Interval, respectively. Though well-established, these approaches impose temporal uniformity that underutilizes human sensitivity to changes in tempo, potentially limiting the perception of derivative features like slope and acceleration. 

Motivated by natural tempo perception \cite{ross2022time} and well-regarded tempo-based sonification systems such as the Geiger counter and ECG \cite{hermann2011sonification}, we introduce Variable Tempo Sampling, a novel sonification method that inverts the sampling logic: data are sampled at uniform y-spacing, leading to notes with consistent pitch spacing played at variable temporal intervals. As slope increases, notes arrive more quickly; when slope decreases, notes are spaced further apart, allowing tempo to act as a perceptual cue for rate-of-change.

The aims for this work are two-fold: 1) to introduce and evaluate Variable Tempo, a novel sampling designed to make slope and acceleration more perceptually salient in univariate functions; and 2) to understand and model how well people compare slopes and discriminate acceleration using Continuous, Variable Pitch Interval, and the newly introduced Variable Tempo methods---along with the data characteristics and user factors that influence performance.

% \begin{enumerate}
%     \item Introduce and evaluate Variable Tempo, a novel sampling designed to make slope and acceleration more perceptually salient in functional relationships.
%     \item Understand and model how well people compare slopes and discriminate acceleration using Continuous, Variable Pitch Interval, and the newly introduced Variable Tempo methods-- along with the data characteristics and user factors that influence performance.
% \end{enumerate}

To investigate these questions, we conducted two psychoacoustic experiments in which sighted participants compared the three sonification strategies: Continuous, Variable Pitch Interval, and Variable Tempo. Experiment 1 (\cref{sec:exp1}) focused on slope ratio magnitude estimation, which quantifies the perceived steepness of a graph’s slope relative to a baseline. This type of judgment is relevant in contexts involving the estimation or comparison of slopes in a graph---for example, estimating how much more rapidly COVID infections increased during the Delta variant surge in the summer of 2021 compared to the Alpha variant wave in the winter of 2020. Experiment 2 (\cref{sec:exp2}) examines acceleration discrimination, or the minimum threshold at which acceleration becomes perceptually identifiable in non-linear data. This type of judgment applies to scenarios where determining rate changes is important, such as whether financial trends are accelerating exponentially or beginning to taper. 

Across both tasks, Variable Tempo consistently outperformed the other methods. In slope ratio magnitude estimation, it yielded greater accuracy, especially for larger slope ratios, along with greater precision. In acceleration discrimination, just-noticeable differences (JNDs) under Variable Tempo were over 13 times smaller than in other conditions. Participants also commonly rated Variable Tempo as easier, more intuitive, less mentally effortful, and preferred.
% This work contributes:

% \begin{itemize}[noitemsep, topsep=0pt]
%     \item Quantitative models of slope and acceleration perception under different pitch-based sonification schemes.
%     \item Variable Tempo Sampling, a novel, perceptually-motivated method that improves sensitivity to rate-of-change.
%     \item Evidence that tempo variation can enhance sonification, making derivative features more perceptible and intuitive.
% \end{itemize}

% These findings highlight how perceptual understanding can guide the design of sonification techniques that align with human auditory strengths.

% Among the auditory dimensions used in sonification, pitch is one of the most common encodings, often representing data magnitude or slope through frequency changes over time \cite{sonification_handbook_chap4_perception, dubus_systematic_2013}. Prior work has investigated how effectiveness of sonification for supporting magnitude estimation, data exploration, polarity, trend identification, trend recreation. However, despite their widespread use \cite{DesmosTone, AppleAudioGraphs, SASGraphicsAccelerator}, we still lack a detailed understanding of how well pitch-based sonification supports the perception of key bivariate data features like slope and acceleration \cite{}. Unlike vision, where psychophysical research has deeply characterized sensitivity to gradients and curvature \cite{}, auditory slope and acceleration perception remains under-modeled \cite{}. 

\section{Related Work}
\label{sec:related-work}

\subsection{Applications of Sonification}
\label{subsect:sonification-applications}

Sonification spans diverse domains, including accessibility, data analysis, education, health, and the arts, both as a standalone technique and as part of multimodal systems \cite{enge2022towards, enge2024open}. For accessibility, it enables users who are blind or have low vision (BLV) to explore scientific data such as NASA imagery \cite{noel2022accessibility} and chemical spectra \cite{Pereira_2013}. In data analysis, sonification leverages auditory sensitivity to patterns in noise \cite{moore2012chapter3, Scaletti_2024}; for instance, non-experts accurately detected epileptic seizures from sonified EEGs \cite{Parvizi_2018}, a task requiring expert interpretation visually. In education and outreach, especially in astronomy \cite{zanella2022sonification, Harrison_2022}, chemistry \cite{Mitchell_2020, smith2024interactive, Scaletti_2022}, and biology \cite{Braun_2024}, it improves engagement, particularly when musical elements are used \cite{middleton_data--music_2023}. In health, sonification shows potential to aid rehabilitation by providing real-time auditory feedback for stroke recovery \cite{scholz_sonification_2016, kantan2023}. In the arts, composers use sonification as a creative tool for transforming data into expressive musical works \cite{Vickers_2016}.

%see also Cherry, 1953 for detection of speech). 

% (98\% in students and 95\% in nurses) compared to experts or non-experts analysing the same information presented visually (88\% in neurologists, 76\% in students).
%see also Cherry, 1953 for detection of speech). 
% Recently, sonifications of protein folding dynamics revealed the role of hydrogen bonding heterogeneity in protein folding transition state passage \cite{Scaletti_2024}.

%As a relatively new and underutilized technique that many in the public are not familiar with, sonification can pique public interest in a topic by connecting it to sound. 
% Analysis of Sonifications of NASA images from 3,184 participants showed significant self-reported learning improvements and positive experiences, in sighted, blind, and low-vision survey participants \cite{Arcand_2024}. Sonifications have been employed in the interpretation of chemical spectra by BVI students, with one study showing that students preferred these methods to Braille\cite{Pereira_2013}.

%These novel approaches supplement already ubiquitous sonification applications in hospitals, such as auditory displays of EKG data. 

\subsection{Sonification Encodings}
\label{subsect:sonification-encodings}

A widely cited definition of sonification is “the transformation of data relations into perceived relations in an acoustic signal for the purposes of facilitating communication or interpretation” \cite{kramer1999sonification}. This emphasis on perceived relations underscores the importance of mapping data onto perceptual dimensions of sound---most commonly pitch, loudness, timbre, and spatialization.

Among these, pitch is the most widely used  \cite{dubus_systematic_2013, hermann2011sonification}. Loudness, while common in auditory displays, is generally considered suboptimal for representing continuous data due to coarse discrimination thresholds, weak auditory memory, and susceptibility to background noise and playback variability \cite{hermann2011sonification, clement1999memory, Flowers_2005}. Timbre---the quality that distinguishes two sounds of identical pitch and loudness---can support categorical encoding in sonification, leveraging the auditory system’s sensitivity to spectral characteristics \cite{bregman1990simultaneous}. Finally, spatialization (e.g., stereo panning or binaural rendering) has been used to supplement pitch and other cues in interactive sonification systems \cite{ronnberg2019interactive, Bujacz2016}, though its utility for continuous data representation remains exploratory.

This study focuses on pitch as the primary encoding dimension due to its prevalence in research \cite{dubus_systematic_2013}, adoption in applied sonification systems \cite{AppleAudioGraphs}, and its high perceptual resolution \cite{kramer1999sonification}.

\subsection{Pitch-based Sonification}
%psychoacoustic work done, and research gaps}
\label{subsect:pitch-encodings}

%Definition of pitch
Pitch is related to frequency logarithmically and measured in units of semitones: $n = 12\times log_2({\frac{f}{440}}) + 69$, where $f$ is frequency in $Hz$ and $n$ is pitch in semitones, with 69 representing $A4$ on the piano.

Human pitch perception is approximately logarithmic in nature, and equal pitch intervals are typically perceived as equidistant, though some evidence suggests compression at higher frequencies \cite{stevens1937, beck1961scaling}.

Pitch sonification leverages this perceptual regularity to encode continuous data trends \cite{Flowers_2005}. To enhance interpretability, pitches are often quantized to musically familiar structures---such as equal-tempered or diatonic scales---which draw on listeners' existing cognitive frameworks \cite{hermann2011sonification}. Even untrained listeners can reliably compare pitch intervals, despite lacking formal terminology \cite{Thompson2012, attneave1971pitch, dowling1971contour, schellenberg1996natural}.

Cultural metaphors further inform pitch mappings. In most languages, pitch is conceptualized vertically---high pitch is associated with higher values or upward motion, and low pitch with lower values or downward motion \cite{pirhonen2008sonification}. Accordingly, sonification systems commonly employ positive polarity mappings. However, in specific contexts (e.g., size or weight), negative polarity mappings---where larger values correspond to lower pitches---may better align with expectations \cite{walker2002magnitude}.

Despite the widespread use of pitch to convey trends, auditory perception of slope and acceleration has received limited empirical attention. 

% and under-modeled—particularly in contrast to the well-characterized sensitivity to gradients in visual perception \cite{talbot2012empirical}.

% In fact, some languages like Indonesian use "small" to describe high pitches (large numerical values) and "large" to describe low pitches (small numerical values).

% Pitch is ubiquitous in sonifications, perhaps the most frequently used perceptual dimension in auditory displays \cite{sonification_handbook_chap4_perception}.
% namely musical scales for sequentially-presented pitches and musical chords for simultaneously-presented ones, 
% as described in equation \ref{eqn:freq-to-pitch}
% Such a mapping is employed, for example, in sonifications of weather data where lower temperature is represented with lower pitch, and higher temperature with higher pitch \cite{flowers2001}. 
%Pitch in sonifications
% The mel scale, proposed by Stevens et al. \cite{stevens1937}, captures the perceptual distance between pitches. In their study, the pitch in mels of other frequencies was determined by asking subjects to adjust a comparison tone until it was perceived to be one half of the pitch height of a standard tone (method of fractionation). Their results indicated that the mel scale and the logarithmic scale are roughly equivalent below 500 Hz, but the mel scale increases at a slower rate above 500 Hz. In other words, perceptually equivalent pitch interval sizes (in mels) span progressively smaller frequency ratios with higher and higher transpositions (\cite{stevens1937, beck1961scaling}).\\

\subsection{Tempo Encoding Explorations}
\label{subsec:tempo-encoding}

Tempo remains relatively underexplored \cite{dubus_systematic_2013} despite the auditory system's remarkable temporal resolution and resilience to noise \cite{POIRIERQUINOT20172}. Early work by Martin \cite{martin1972rhythmic} framed rhythm---the temporal structuring of sound events---as central to auditory information processing. He argued that our attentional system is attuned to anticipate rhythmic events in many serial stimuli that include speech and music. Stimuli that align with listeners' temporal expectations are often more perceptually salient and easier to process.

Numerical values can be mapped to temporal parameters such as duration or rate. For example, sound stream duration has been successfully used in auditory histograms, while temporal patterning can communicate changes in distributional properties over time \cite{Flowers_2005}. Tempo and rhythm can also provide avenues for creating temporally engaging structures \cite{smith2024sonifications} and can enhance user enjoyment, reduce fatigue, and promote sustained engagement \cite{sonification_handbook_chap7_aesthetics, coers2023movement}.

Musical training may further modulate listeners’ sensitivity to tempo-based cues. Prior studies show that musicians exhibit lower thresholds for both pitch and tempo discrimination, greater attentional acuity for non-speech sounds, and enhanced perceptual acuity overall \cite{kishon-rabin_pitch_2001, marie2012musical, janata_acuity_2006}. As such, considering listener expertise is important in evaluating the effectiveness of different auditory mappings.

We explore tempo not as an explicit encoding of data values, but as an emergent property of our sampling method---specifically, sampling data at equal y-intervals rather than equal x-intervals. This produces sonifications where pitch intervals remain fixed and timing varies in proportion to the data’s slope. In this way, tempo becomes an implicit perceptual cue for interpreting slope and acceleration.

%The importance of aesthetic considerations in sonifications (and visualizations) has been acknowledged \cite{tractinsky2004toward}
%\cite{norman2004emotional}
% \cite{kramer1994organizing}\cite{grond2014interactive}, 
%at minimum, proper aesthetic considerations ensures 

\section{Sampling Approaches and Study Conditions}

\begin{figure*}[h]% specify a combination of t, b, p, or h for top, bottom, on its own page, or here
  \centering % avoid the use of \begin{center}...\end{center} and use \centering instead (more compact)
  \includegraphics[width=2\columnwidth, alt={Diagram illustrating three sonification strategies for data functions: Continuous, Variable Pitch Interval (equal x-spacing), and Variable Tempo (equal y-spacing). Section A shows a smooth, sublinearly growing curve labeled "y = f(x)". Section B shows three data sampling methods: Continuous, which samples data at infinitesimally small intervals (dx), Equal x-spacing, which samples data at regular horizontal intervals (delta x), and Equal y-spacing, which samples data so that each step changes the y-value by approximately delta y, with corresponding x-values estimated. Section C shows encoding. In all methods, y is mapped to pitch using a logarithmic function, and x is mapped to time. Continuous sampling produces a continuous note, while the other two produce discrete notes. Section D shows the resulting sonifications for the sublinearly growing example curve. Continuous sampling results in a smooth pitch glide over time. Equal x-spacing produces evenly timed notes with progressively smaller pitch intervals (Variable Pitch Interval). Equal y-spacing produces evenly spaced pitches with progressively longer temporal intervals (Variable Tempo). Section E shows an example Experiment 1 trial of two affine trends with different intercepts and slopes. Participants estimate which trend is steeper and by what factor. Section F shows an example Experiment 2 trial of a curve that is accelerating upwards. Participants judge whether the curvature is upwards or downwards.}]{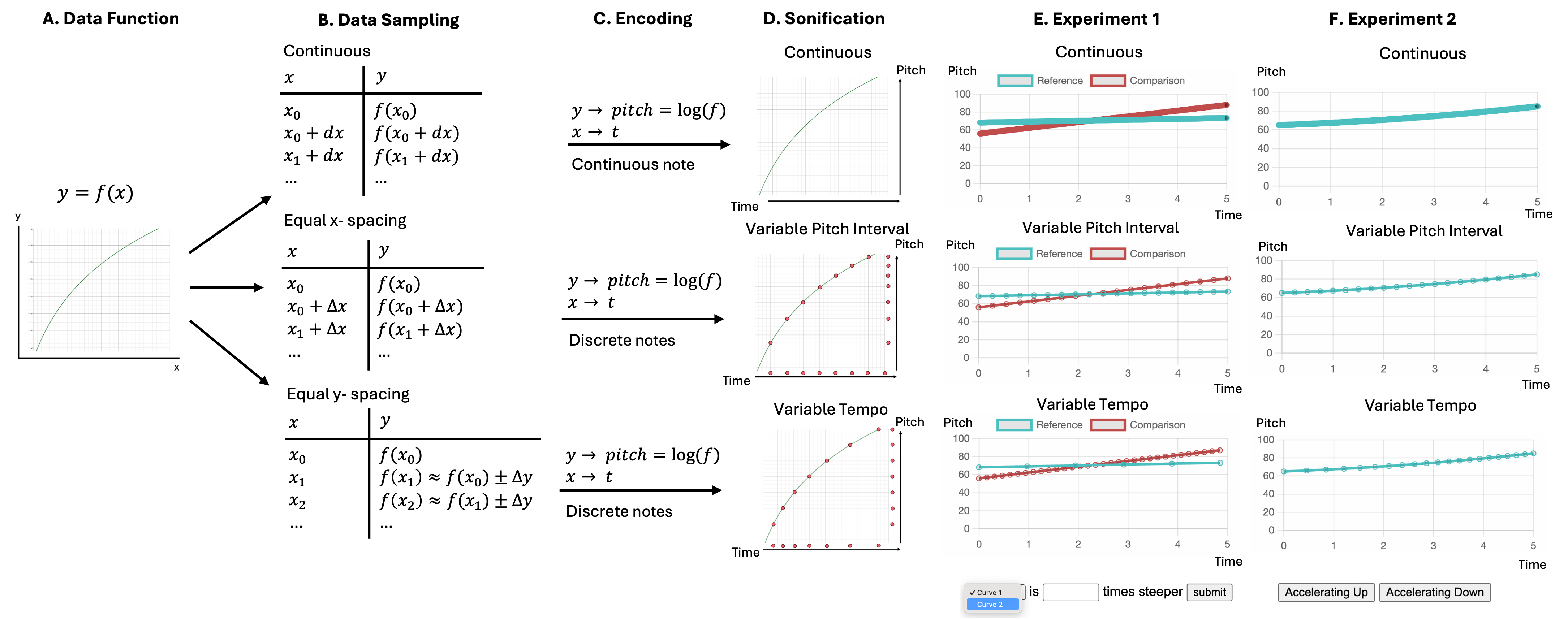}
  
  \caption{(A) A continuous data function y=f(x) serves as the underlying phenomenon to be sonified. (B) Three data sampling strategies are shown: continuous sampling, equal x-spacing (used in Variable Pitch Interval), and equal y-spacing (used in Variable Tempo). Continuous sampling reflects the raw function as-is, while equal x- and y-spacing discretize the function based on consistent increments in either the horizontal or vertical axis, respectively. (C) All strategies map y to pitch and encode x as time, but differ in how notes are sampled. (D) Resulting sonifications show how each method produces different temporal distributions of notes: Continuous generates a smooth pitch glide; Variable Pitch Interval uses equal time steps with differently spaced pitch values; and Variable Tempo keeps pitch intervals constant but varies timing between notes to reflect data slope. (E) In Experiment 1, participants listened to two randomly sampled clips with different slopes and intercepts and estimated which slope was steeper and by what factor. (F) In Experiment 2, participants listened to curves with positive or negative acceleration and estimated the acceleration direction.}
  \label{fig:samplingIntro}
\end{figure*}

This work compares three sonification sampling methods---Continuous, Variable Pitch Interval, and Variable Tempo (\cref{fig:samplingIntro})---used for continuous univariate functions. Each method maps the same underlying data function to a sequence of pitches over time using a 2D auditory mapping: pitch encodes the dependent variable, and time encodes the independent variable. Although the overall pitch trajectory is determined by the same underlying function, the sampling strategy affects both which pitches are played and when they are played, altering the temporal structure of the display and potentially influencing perception.

\subsection{Continuous}

In Continuous sampling, data are sonified as a smooth, uninterrupted curve, producing a continuous pitch glide reflecting the data’s shape (\cref{fig:samplingIntro} Top). This mirrors common auditory graphing tools and serves as a baseline for comparison \cite{AppleAudioGraphs, hermann2011sonification}. Estimating slope in this condition requires assessing the rate of pitch change, or $d(pitch)/dt$.

\subsection{Variable Pitch Interval}

Variable Pitch Interval sampling (or equal-x sampling) follows a standard sonification approach: data are sampled at uniform x-intervals, with each y-value mapped to pitch \cite{hermann2011sonification} (\cref{fig:samplingIntro} Middle). This yields evenly spaced notes where pitch changes reflect the slope at each point. With $\Delta t$ held constant, listeners rely on $\Delta pitch$ to judge slope---steeper regions produce larger pitch intervals, flatter ones smaller.

\subsection{Variable Tempo}
\label{sec:VariableTempo}

This work introduces Variable Tempo sampling, which reverses the conventional sampling logic: data are sampled at uniform increments along the y-axis rather than the x-axis (\cref{fig:samplingIntro} Bottom). This produces a sequence of constant pitch intervals ($\Delta pitch$) with variable note timing ($\Delta t$) that is faster in steeper regions and slower in flatter ones.

For accelerating curves that can be modeled or locally approximated as $pitch(t) = a*t^2 + b*t + c$, the instantaneous slope is $\frac{d(pitch)}{dt} = 2*a*t + b$.

To maintain constant $\Delta pitch$ (in which $\Delta pitch$ is positive for increasing slopes and negative for decreasing slopes), the time between notes must vary inversely with slope: $\Delta t = \frac{\Delta pitch}{2*a*t + b}$, yielding notes that play at frequency $freq(t) = \frac{1}{\Delta t} = \frac{2*a*t}{\Delta pitch} + \frac{b}{\Delta pitch}$.

Initial tempo is the ratio of starting slope ($b$) and pitch increment ($\Delta pitch$), and increases at a rate dictated by the ratio of acceleration ($a$) and pitch increment ($\Delta pitch$). For the example of an inverse parabola $y = -x^2 + 2x$ (mapping to $pitch(t) = -t^2 + 2t$), the local tempo starts fast (dominated by $\frac{2}{\Delta pitch}$), slows to zero at the inflection point (where $\frac{2*(-1)*t}{\Delta pitch}$ = $-\frac{2}{\Delta pitch}$), and then increases again beyond the inflection point (now dominated by $\frac{2*(-1)*t}{\Delta pitch}$).

\section{Experiment 1: Slope Ratio Magnitude Estimation}
\label{sec:exp1}

Slope comparison involves the estimation of how much steeper or shallower one slope is relative to another. This can be quantified as a slope ratio magnitude ($|SR| = |slope_{comp} / slope_{ref}|$). Experiment 1 evaluated slope ratio magnitude estimation under the three sonification sampling methods: Continuous, Variable Pitch Interval, and Variable Tempo. Given two sloped pitch functions of the form $pitch(t) = b \cdot t + c$ (where $b$ is the slope in real-valued $semitones/s$ and $c$ is the intercept in $semitones$), we examined how slope ratio magnitude estimation accuracy and precision between the two functions were influenced by (1) slope ratio magnitude ($\log(|SR|)$), (2) slope sign agreement ($\text{sign}(SR)$), and (3) participants’ musical training ($\text{MusicExp}$). We also collected subjective ratings of confidence and mental effort, along with qualitative feedback on strategies, difficulty, and method preference.

% (Table~\ref{tab:model_variables}). 

% \begin{table}[h]
%   \caption{%
%   	Experiment \ref{sec:exp1} model variables.%
%   }
%   \label{tab:model_variables}
%   \scriptsize%
%   \centering%
%   \begin{tabu}{%
%     >{\raggedright\arraybackslash}p{0.12\linewidth}%
%     >{\raggedright\arraybackslash}p{0.20\linewidth}%
%     >{\raggedright\arraybackslash}p{0.54\linewidth}%
%   }
%   \toprule
%   \textbf{Variable Name} & \textbf{Parameter Name} & \textbf{Description} \\
%   \midrule
%   \texttt{Condition} & Condition & Sonification sampling method. \\
%   \addlinespace
%   \texttt{log(|SR|)} & Log Slope Magnitude Ratio & Log-transformed ratio between reference (S1) and comparison (S2) slopes to promote symmetry between S1$>$S2 and S2$<$S1 cases. \\
%   \addlinespace
%   \texttt{sign(SR)} & Slope Sign Agreement & Indicator for whether reference and comparison slopes have the same or opposite direction. \\
%   \addlinespace
%   \texttt{MusicExp} & Years of music training & Self-reported duration of music training. \\
%   \bottomrule
%   \end{tabu}%
% \end{table}

\subsection{Study Procedure and Stimuli}
\label{sec:exp1procedure}

Participants were prescreened via an online survey, which excluded those with self-reported hearing loss or incorrect responses to two graphical screening questions comparing slope steepness. The survey also collected informed consent and demographic information, including age range, gender, data and statistical background, sonification familiarity, and years of formal music training. Detailed participant demographics are provided in the Supplemental Material.

Participants wore Audio-Technica ATH-M50x over-ear headphones during an hour-long, in-person session. The study consisted of a calibration phase, followed by the experimental conditions---Continuous, Variable Pitch Interval, and Variable Tempo---each comprising training, practice, the main task, and a post-task interview. The ordering of conditions was counterbalanced across participants. Each participant received a 30USD gift card for their participation. All procedures were approved by the Institutional Review Board.

During calibration, a facilitator played the highest (MIDI 88) and lowest (MIDI 55) pitches used in the study and adjusted the volume for comfort and clarity.

During training, facilitators provided a brief explanation of the mapping strategy, followed by a session featuring four example slopes---two increasing and two decreasing---varying in steepness. Participants were encouraged to think aloud and ask questions as they identified the direction and relative steepness of each slope.

Practice consisted of five example trials in which participants listened to two randomly sampled audio clips with different slopes and intercepts. Afterward, they answered: (1) “Which one is steeper?” and (2) “By what factor?” Feedback was provided after each trial to help calibrate their responses.

The main task comprised 25 trials per condition, each involving a single exposure to the pair of audio clips. In each trial, participants judged which of two randomly sampled audio clips---a reference and a comparison slope---was steeper and by what factor. Each stimulus lasted exactly 5 seconds, with fade-in and fade-out ramps to discourage note counting and promote attention to the rate of change \cite{Loomis1998Assessing}. In the Variable Tempo condition, notes were triggered at every semitone change. To match note counts across conditions, x-spacings in Variable Pitch Interval were chosen to produce the same number of notes as the corresponding trend in Variable Tempo.

After each condition block, participants briefly described their strategy and rated (1) their confidence and (2) perceived mental effort on a 7-point Likert scale (1 = low, 7 = high). Upon completing all three conditions, they reflected on which condition they found most challenging, most accurate, and most preferred.

To span a broad and balanced range of slope ratios and directions, we sampled slope ratios uniformly in log space over the range $\pm \log((88 - 55)/5)$. Reference and comparison slopes were sampled to match the sampled slope ratio, constrained to viable bounds: both slopes fell between $\pm(88 - 55)/5$ semitones per second and exceeded 1 semitone/s in magnitude. Intercepts were randomly chosen such that the full 5-second pitch trajectory stayed within the MIDI 55–88 range for perceptual clarity and comfort \cite{brown2003drawing, wang2022seeing}.

To reduce pitch-loudness confounds, sounds were loudness-normalized using the ISO 226 40-phon contour \cite{suzuki2024}. Sonifications were synthesized with sine oscillators for timbral simplicity and rendered via a custom WebChuck application \cite{chafe2023would}.

\subsection{Participants}

Twelve sighted individuals (10 women, 1 non-binary, 1 self-described), aged 18–34, participated in the study. Participants were recruited via large university mailing lists. All were prescreened to confirm a baseline understanding of slopes and no self-reported hearing loss.

Five participants reported prior exposure to sonification, though interaction was minimal (e.g., “never” or “fewer than once a year”). Most (9/12) had some formal musical training: two reported less than 1 year, one had 1–2 years, four had 2–5 years, and two had 5–10 years. None had absolute pitch. Four had completed graduate-level coursework or professional training in data analysis, seven had undergraduate-level coursework, and one had K–12-level exposure.

\subsection{Analysis}

\textbf{Estimation Performance:} We modeled estimation error using a Bayesian hierarchical framework to assess how sampling method, stimulus-level features, and participant-level features influence slope ratio magnitude estimation, including their interactions with condition \cite{Gelman2013philosophy, Gelman2003bayesian}. Partial pooling across participants was supported through random intercepts, and residual variability was explicitly modeled by including the same fixed and random effects in the sigma component.

Accuracy was measured as the mean of ln–transformed error, and precision as the residual standard deviation. Ln-transformed error---defined as $\ln(\hat{SR}/SR)$ where $\hat{SR}$ is the estimated slope ratio and $SR$ the true value. Unlike percent error, which yields asymmetric values for equivalent over- and underestimates (e.g., $+100\%$ vs. $-50\%$ for a twofold overestimate of larger slope), ln estimation error yields $\pm \ln(2)$ in both cases, providing an unbiased, direction-independent evaluation regardless of which slope is steeper or presented first. As with percent error, $ln(\hat{SR} / SR) = 0$ indicates a perfectly accurate estimate, while positive and negative values indicate overestimation and underestimation, respectively. At small error magnitudes, ln estimation error closely approximates percent error (e.g. $ln(1.05) \approx 4.9\%$), but diverges at larger values.

We used a skew-normal likelihood to account for the asymmetric residual distributions observed during initial visual checks. Weakly informative priors were applied to fixed effects ($\mathcal{N}(0, 5)$). Posterior estimates were stable across three prior specifications (between $\mathcal{N}(0, 5)$ to $\mathcal{N}(0, 1)$), with maximum absolute differences in posterior means, medians, standard deviations, and interquartile ranges all below 0.001. The model structure was as follows: $\text{LnError} \sim 1 + \text{Condition} \times (\text{log(|SR|)} + \text{sign(SR)} + \text{MusicExp}) + (1 | \text{Participant})$; $\text{sigma} \sim 1 + \text{Condition} \times (\text{|log(|SR|)|} + \text{sign(SR)} + \text{MusicExp)} + (1 | \text{Participant)}$

All MCMC chains converged ($\hat{R} \approx 1.00$), with no divergent transitions or sampling issues. Posterior predictive checks indicated good model fit, with predictive p-values of $0.368$ (median), $0.317$ (mean), $0.703$ (IQR), and $0.506$ (SD), suggesting the model captured both central tendency and dispersion well. Modeling was conducted using the brms package in R \cite{burkner2017brms}. Models were fit using Hamiltonian Monte Carlo (HMC) with max\_treedepth = 25 and adapt\_delta = 0.98 to improve convergence stability.

\textbf{Participant Ratings:} To evaluate condition effects on confidence and mental effort ratings, we fit Bayesian ordinal regression models using a cumulative logit link and weakly informative priors ($\mathcal{N}(0, 5)$): $\text{Confidence/Mental Effort} \sim \text{Condition} + (1 | \text{Participant)}$

\textbf{Strategies and Reflection:} Participants’ open-ended responses were thematically coded to identify common strategies and perceived differences across conditions.

\subsection{Results}

\subsubsection{Performance}

\begin{figure}[h]% specify a combination of t, b, p, or h for top, bottom, on its own page, or here
  \centering % avoid the use of \begin{center}...\end{center} and use \centering instead (more compact)
  \includegraphics[width=\columnwidth, alt={Violin plot showing ln estimation error distributions across three sonification conditions: Continuous (Cont.), Variable Pitch Interval (Var. Pitch Int.), and Variable Tempo (Var. Tempo). Each violin plot contains individual trial dots (black), a central dot indicating the mean, and a vertical line representing the 95\% confidence interval. A horizontal dashed red line at ln estimation error = 0 marks unbiased estimation. All three plots are centered near ln estimation error = 0. The Variable Tempo condition shows a tighter distribution compared to the other two conditions.}]{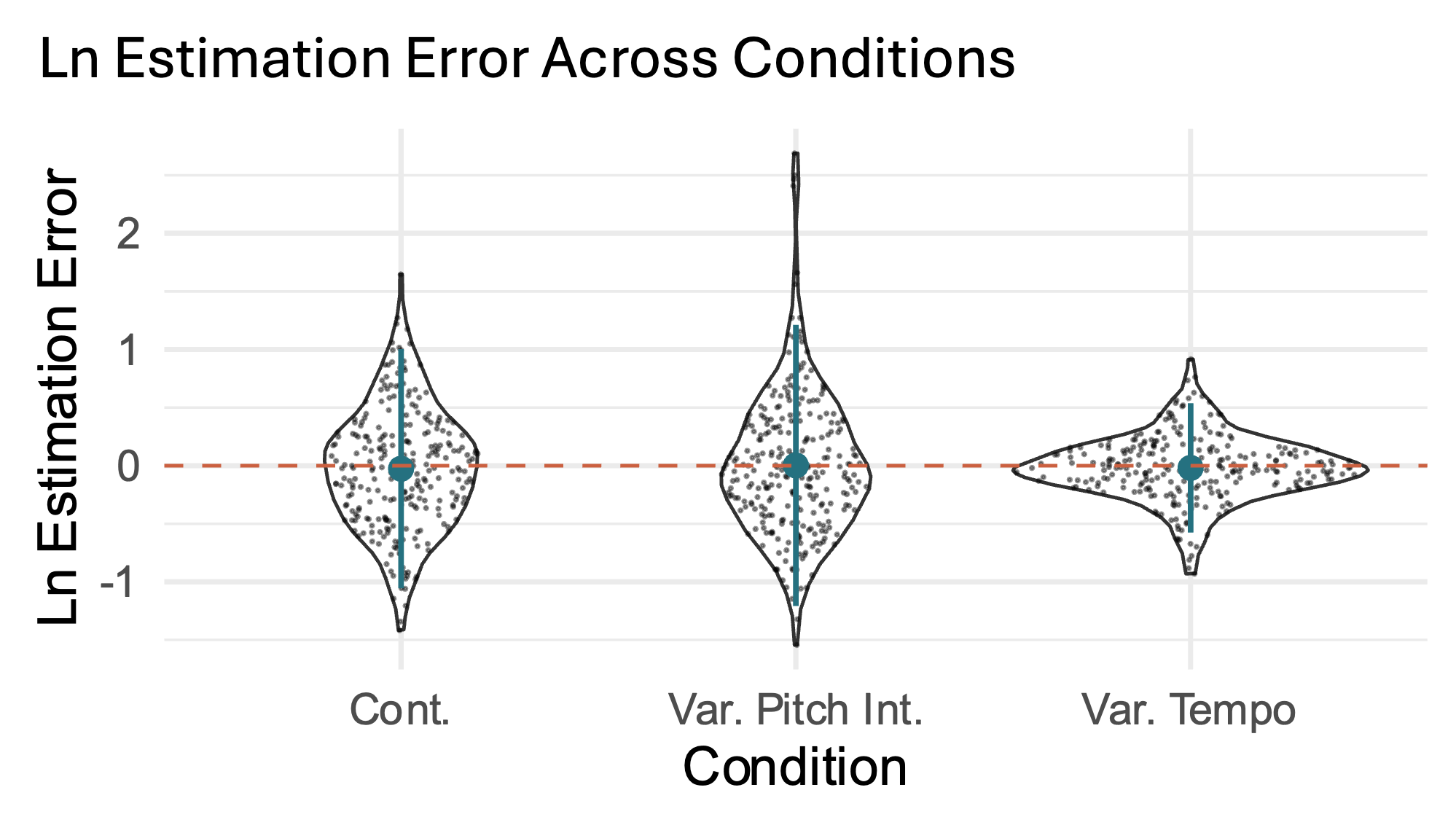}
  \caption{Distribution of participant estimation errors for each condition across all trials. Dots represent individual trials, and central points and bars indicate the mean and $95\%$ confidence interval.}
  \label{fig:part1Main}
\end{figure}

\cref{fig:part1Main} shows the distribution of estimation errors for each condition: Continuous ($\mu = -0.0262, \sigma = 0.514$), Variable Pitch Interval ($\mu = 0.00213, \sigma = 0.605$), and Variable Tempo ($\mu = -0.0192$, $\sigma = 0.278$).

\begin{figure}[h]% specify a combination of t, b, p, or h for top, bottom, on its own page, or here
  \centering % avoid the use of \begin{center}...\end{center} and use \centering instead (more compact)
  \includegraphics[width=\columnwidth, alt={Figure titled "Intercept-Adjusted Estimates for Ln Estimation Error" showing posterior model estimates of Ln Estimation Error for different predictors across three sonification conditions: Continuous, Variable Pitch Interval, and Variable Tempo. Four panels (A to D) display point estimates with 95\% credible intervals for each predictor. (A) Intercept-adjusted ln estimation errors across conditions are –0.02 [CI = –0.11, 0.08] for continuous, –0.08 [CI = –0.17, 0.02]* for Var. Pitch Int., and 0.00 [CI = –0.06, 0.07] for Var. Tempo. Asterisks beside Var. Pitch Int. indicate posterior probability > 95\%. Vertical bar with asterisks shows credible pairwise difference between Var. Pitch Int. and Var. Tempo. (B) Interaction with log slope ratio (log(|SR|)) and Continuous is –0.28 [CI = –0.40, –0.16]***, Var. Pitch Int. is –0.58 [CI = –0.69, –0.47]***, and Var. Tempo is –0.13 [CI = –0.19, –0.07]***. Triple asterisks denote posterior probability > 99.9\% for all estimates. Vertical bars with asterisks indicate credible pairwise differences among all three conditions. (C) Interaction with slope sign agreement (sign(SR)) and Continuous is –0.01 [CI = –0.12, 0.09], Var. Pitch Int. is 0.01 [CI = –0.09, 0.11], and Var. Tempo is –0.03 [CI = –0.09, 0.02]. No estimates show credible effects; all 95\% CIs include zero. (D) Interaction with years of music training (MusicExp) and Continuous is –0.00 [CI = –0.02, 0.01], Var. Pitch Int. is 0.02 [CI = 0.00, 0.03]*, and Var. Tempo is –0.00 [CI = –0.01, 0.01]. An asterisk next to Var. Pitch Int. indicates a posterior probability > 95\%. Vertical bar with asterisks marks credible differences between Var. Pitch Int. and the others.}]{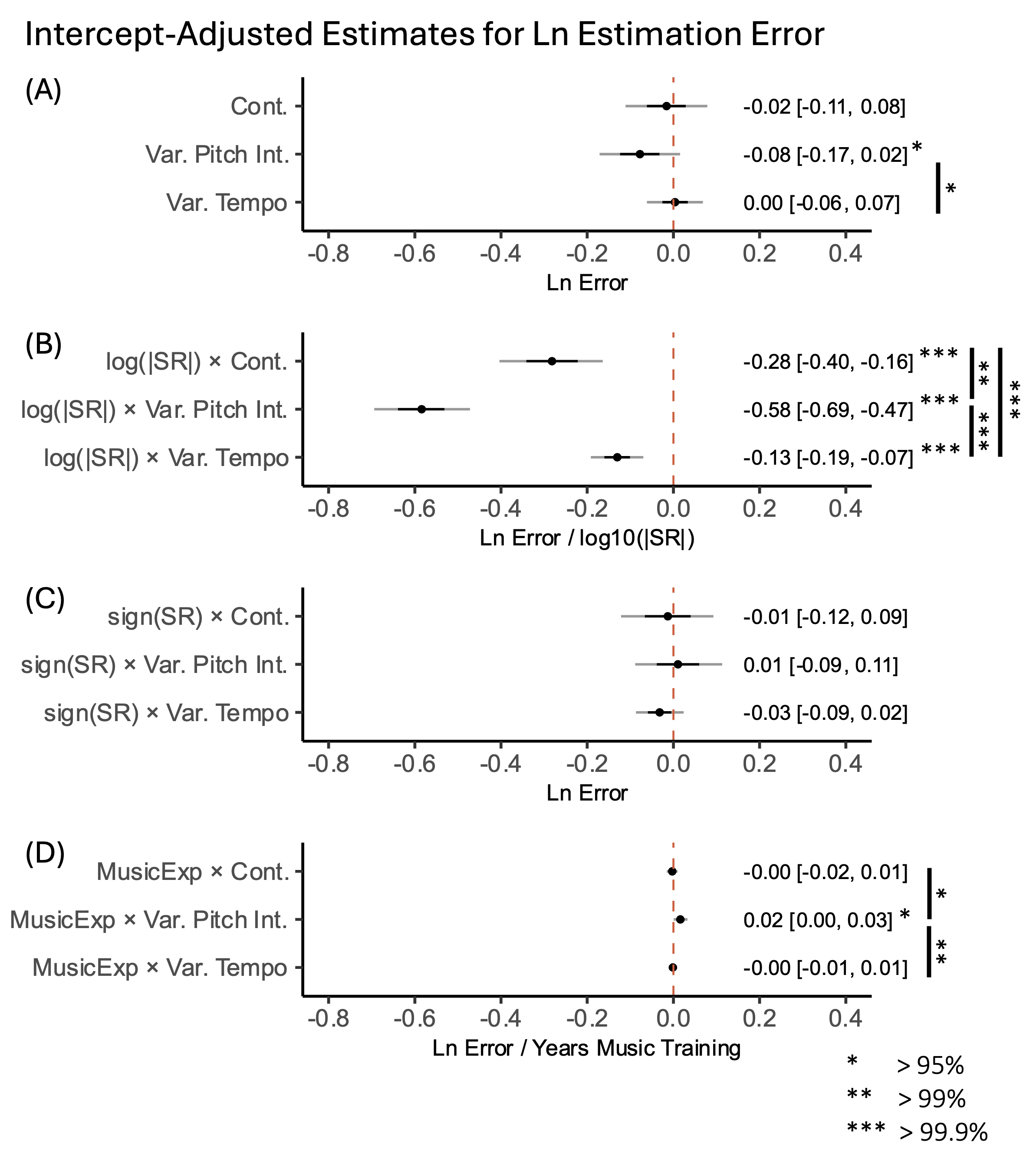}
  \caption{Intercept-adjusted model estimates of ln estimation error contributions across sonification conditions. Posterior means and $95\%$ credible intervals are shown for (A) condition-level intercepts, (B) condition $\times$ slope magnitude ratio interactions, (C) condition $\times$ slope sign agreement interactions, and (D) condition $\times$ musical training interactions. Asterisks denote posterior probability of direction ($** > 99\%$, $*** > 99.9\%$).}
  \label{fig:part1AdjEst}
\end{figure}

\begin{figure}[h]% specify a combination of t, b, p, or h for top, bottom, on its own page, or here
  \centering % avoid the use of \begin{center}...\end{center} and use \centering instead (more compact)
  \includegraphics[width=\columnwidth, alt={Scatterplot titled "Ln Estimation Error Across |Slope Ratio| and Condition" showing ln estimation error plotted against the absolute value of slope ratio (|SR|) across three sonification conditions: Continuous, Variable Pitch Interval, and Variable Tempo. Each subplot contains individual participant trial data as black dots and a blue trend line representing the fitted regression line for each condition. The x-axis (shared across plots) is |SR|, ranging from approximately 0.2 to 5.0 on a logarithmic scale. The y-axis is ln estimation error, ranging from approximately -1.5 to 2.5. A red dashed horizontal line at y = 0 marks the reference for unbiased estimation. In the Continuous and Variable Pitch Interval conditions, ln estimation error decreases as |SR| increases, indicating growing underestimation of steeper slopes. In the Variable Tempo condition, the regression line is flatter, suggesting less bias across |SR| values.}]{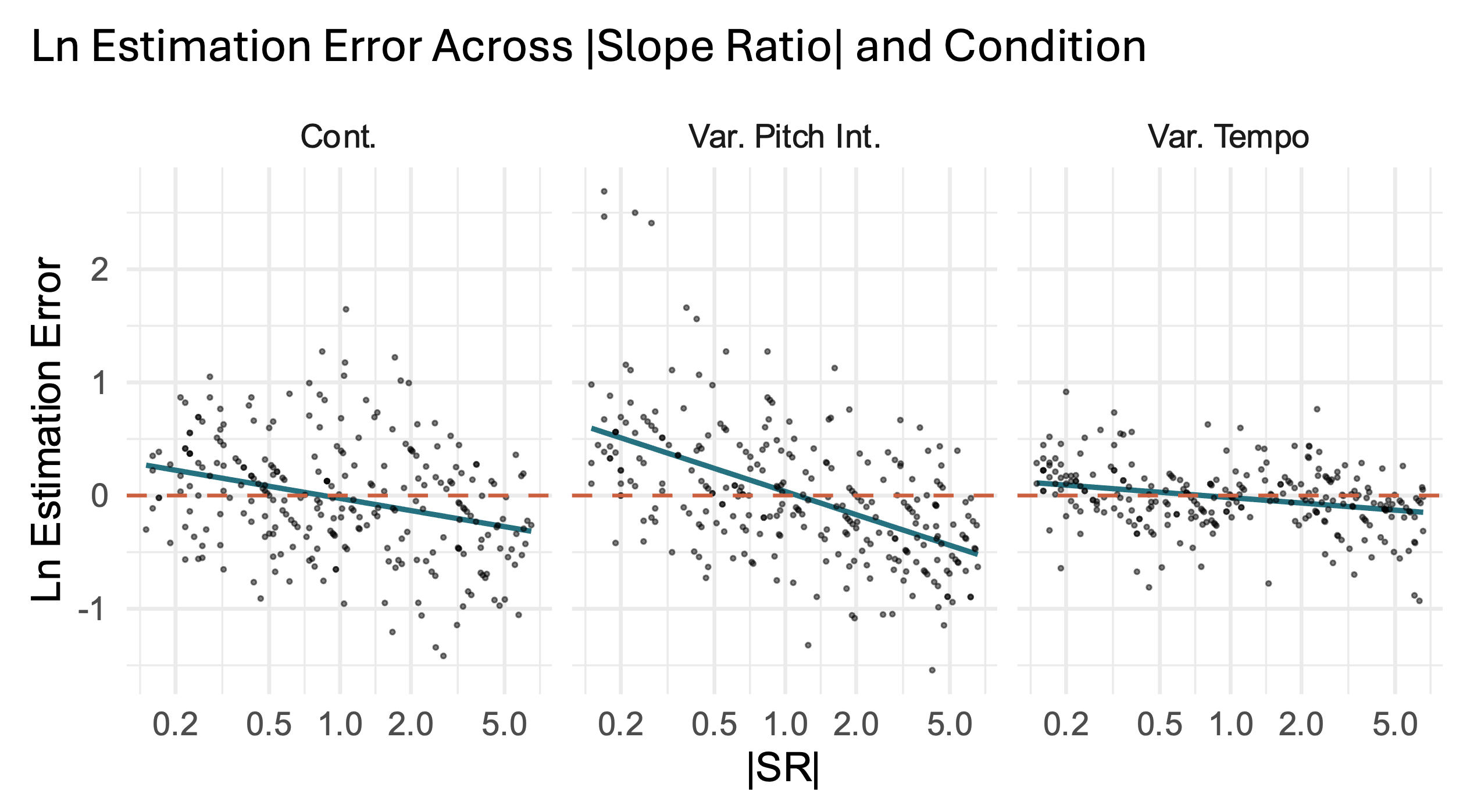}
  \caption{Ln estimation errors across conditions as a function of slope ratio magnitude. Each point represents a single trial’s ln estimation error. The x-axis is plotted on a logarithmic scale to reflect the perceptual scaling of slope ratios. Solid lines indicate condition-specific regression fits.}
  \label{fig:part1SR}
\end{figure}

\textbf{Accuracy:} At the baseline slope ratio ($SR = 1$, $slope_{comp} = slope_{ref}$), the model identified a slight negative bias in ln estimation error for Variable Pitch Interval ($\beta = -0.08$ [$-0.17$, $0.02$]), indicating a modest tendency to underestimate the comparison slope relative to the reference (\cref{fig:part1AdjEst}A). As slope ratio increased beyond $SR = 1$ ($slope_{comp} > slope_{ref}$), all conditions showed increasingly negative ln estimation error, reflecting systematic underestimation of the steeper comparison slope over the reference (\cref{fig:part1SR}). Conversely, as slope ratio decreased below $SR = 1$ ($slope_{comp} < slope_{ref}$), all conditions showed increasingly positive ln estimation error, reflecting systematic overestimation of the shallower comparison slope over the reference. Taken together, these patterns indicate a consistent tendency to underestimate the steeper of the two slopes, regardless of whether it was the comparison or the reference. This effect was strongest in Variable Pitch Interval ($\beta = -0.58$ [$-0.69$, $-0.47$]), followed by Continuous ($\beta = -0.28$ [$-0.40$, $-0.16$]) and Variable Tempo ($\beta = -0.13$ [$-0.19$, $-0.07$]), with posterior probabilities for all pairwise differences exceeding $99\%$ (\cref{fig:part1AdjEst}B).

Effects of slope sign agreement were small and uncertain, with all $95\%$ credible intervals overlapping zero, suggesting no consistent impact on estimation error (\cref{fig:part1AdjEst}C). Musical training also showed minimal overall influence, though baseline underestimation bias of the comparison slope in the Variable Pitch Interval condition appeared slightly reduced among participants with more musical training (\cref{fig:part1AdjEst}D).

\begin{figure}[h]% specify a combination of t, b, p, or h for top, bottom, on its own page, or here
  \centering % avoid the use of \begin{center}...\end{center} and use \centering instead (more compact)
  \includegraphics[width=\columnwidth, alt={Figure titled "Intercept-Adjusted Back-Transformed Residual (Sigma) Estimates for Ln Estimation Error" showing posterior estimates of residual variability in ln estimation error across sonification conditions. Four panels (A to D) display point estimates and 95\% credible intervals for standard deviation (SD) or SD multipliers. (A) Residual standard deviations (SD) of ln estimation error by condition are 0.43 [0.31, 0.60]*** for Continuous, 0.34 [0.24, 0.46]*** for Var. Pitch Int., and 0.21 [0.15, 0.30]*** for Var. Tempo. Triple asterisks indicate > 99.9\% posterior probability. Asterisks with bars indicate all pairwise differences are >95\%. (B) Effect of slope ratio magnitude (log(|SR|)) on SD (as residual SD multipliers e^β, applied as (e^β)^log(|SR|)) for Continuous is 0.81 [0.53, 1.19], Var. Pitch Int. is 1.13 [0.77, 1.61], and Var. Tempo is 1.12 [0.73, 1.66]. All 95\% credible intervals overlap 1, suggesting uncertainty about influence. (C) Effect of slope sign agreement on SD (for negative slope sign) for Continuous is 1.25 [1.05, 1.49]**, Var. Pitch Int. is 1.26 [1.04, 1.51]*, and Var. Tempo is 0.99 [0.82, 1.18]. Asterisks indicate credible evidence for Continuous and Var. Pitch Int. conditions. Vertical bars show credible pairwise difference between Var. Tempo and the others. (D) Effect of musical training on SD (applied as (e^β)^MusicExp) for Continuous is 1.01 [0.97, 1.05], Var. Pitch Int. is 1.03 [0.99, 1.07], and Var. Tempo is 1.02 [0.98, 1.06]. All estimates are near 1 with overlapping credible intervals, indicating minimal influence.}]{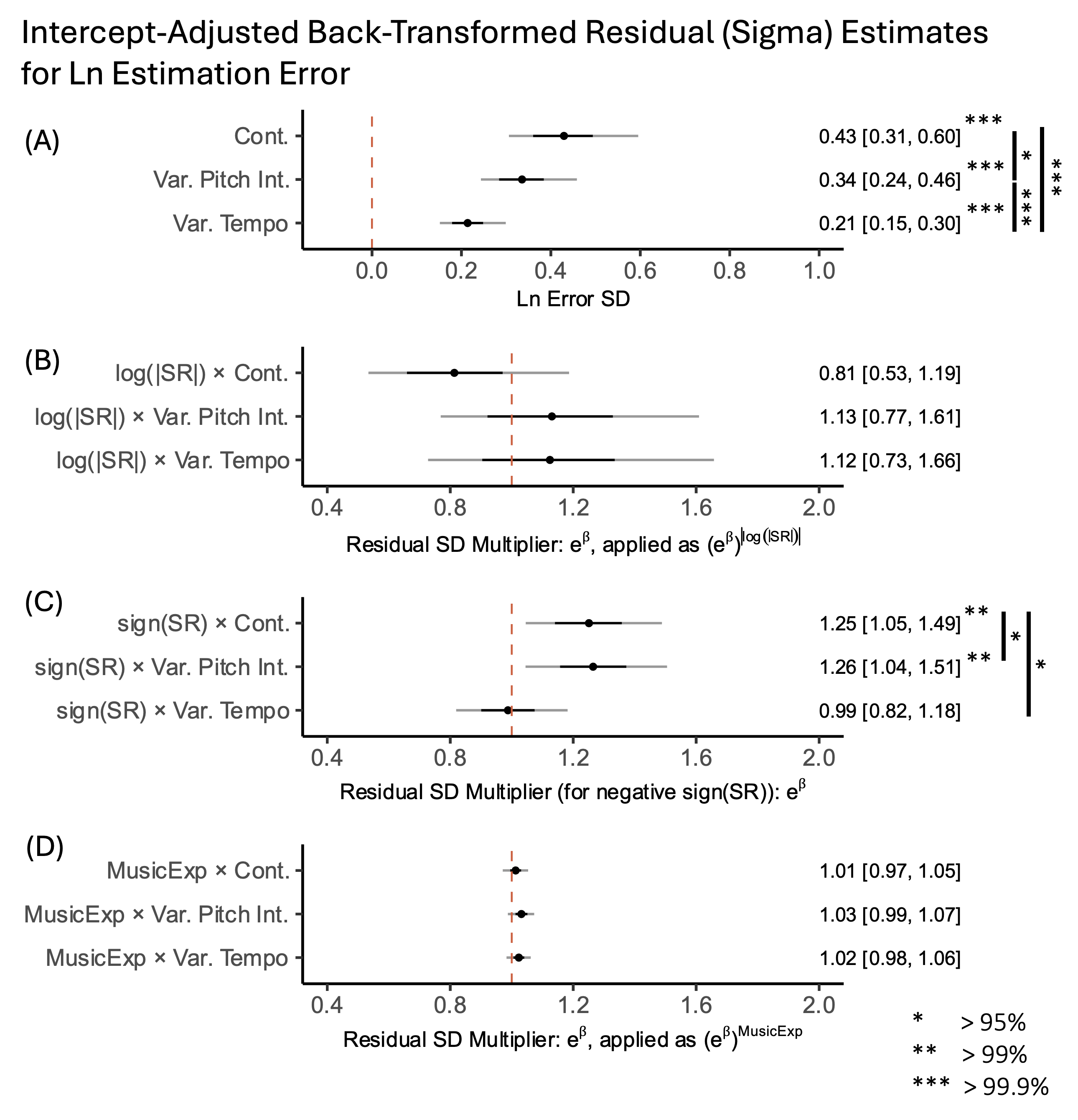}
  \caption{Intercept-adjusted, back-transformed estimates of residual variability (standard deviation) across conditions and predictors. (A) Baseline SDs by condition, adjusted for the intercept. (B--D) Residual SD multipliers expressed as $e^\beta$, where values represent multiplicative changes in SD. Multipliers are applied as $(e^\beta)^x$, where $x$ is (B) $\log(|\text{SR}|)$, (C) $1[\text{sign(SR)} < 0]$, or (D) $\text{MusicExp}$. Values $>1$ indicate increased variability; values $< 1$ indicate decreased variability. Error bars represent $95\%$ (black) and $66\%$ (gray) credible intervals. Asterisks denote posterior probability of direction: $* >95\%$,  $** >99\%$, $*** >99.9\%$.}
  \label{fig:part1AdjSigEst}
\end{figure}

\textbf{Precision:} At the baseline slope ratio ($SR = 1$, $slope_{comp} = slope_{ref}$) and with slope sign agreement ($\text{sign}(SR) = 1$), Variable Tempo resulted in the most consistent judgments with the lowest residual variability ($SD = 0.21\ [0.15, 0.30]$). This was followed by Variable Pitch Interval ($SD = 0.34\ [0.24, 0.46]$) and Continuous ($SD = 0.43\ [0.31, 0.60]$), with all differences strongly supported (\cref{fig:part1AdjSigEst}A). Precision also decreased when participants compared slopes with opposite signs (e.g., a positive vs. negative trend) for both Variable Pitch Interval ($e^\beta = 1.26\ [1.04, 1.51]$) and Continuous ($e^\beta =1.25\ [1.05, 1.49]$) (\cref{fig:part1AdjSigEst}C) but not Variable Tempo ($e^\beta = 0.99\ [0.82, 1.18]$). No credible effects on variability were observed for $|\text{Slope Ratio}|$ (\cref{fig:part1AdjSigEst}B) or musical training (\cref{fig:part1AdjSigEst}D).

\subsubsection{Participant Ratings}

\begin{figure}[h]% specify a combination of t, b, p, or h for top, bottom, on its own page, or here
  \centering % avoid the use of \begin{center}...\end{center} and use \centering instead (more compact)
  \includegraphics[width=\columnwidth, alt={Two line and dot plots showing participant confidence and mental effort ratings across three sonification conditions: Continuous, Variable Pitch Interval, and Variable Tempo. The left plot shows confidence ratings between 1 and 7. Mean confidence scores are higher for Variable Tempo than Variable Pitch Interval, which is higher than Continuous. Colored lines connect individual participant ratings. Asterisks indicate credible difference (>99\%) between all conditions. The right plot shows mental effort ratings between 1 and 7. Mean mental effort scores are slightly higher in Variable Pitch Interval than Continuous, which is slightly higher than Variable Tempo. Colored lines connect individual participant ratings. Asterisks indicate credible differences (>99\%) between Variable Pitch Interval and Variable Tempo.}]{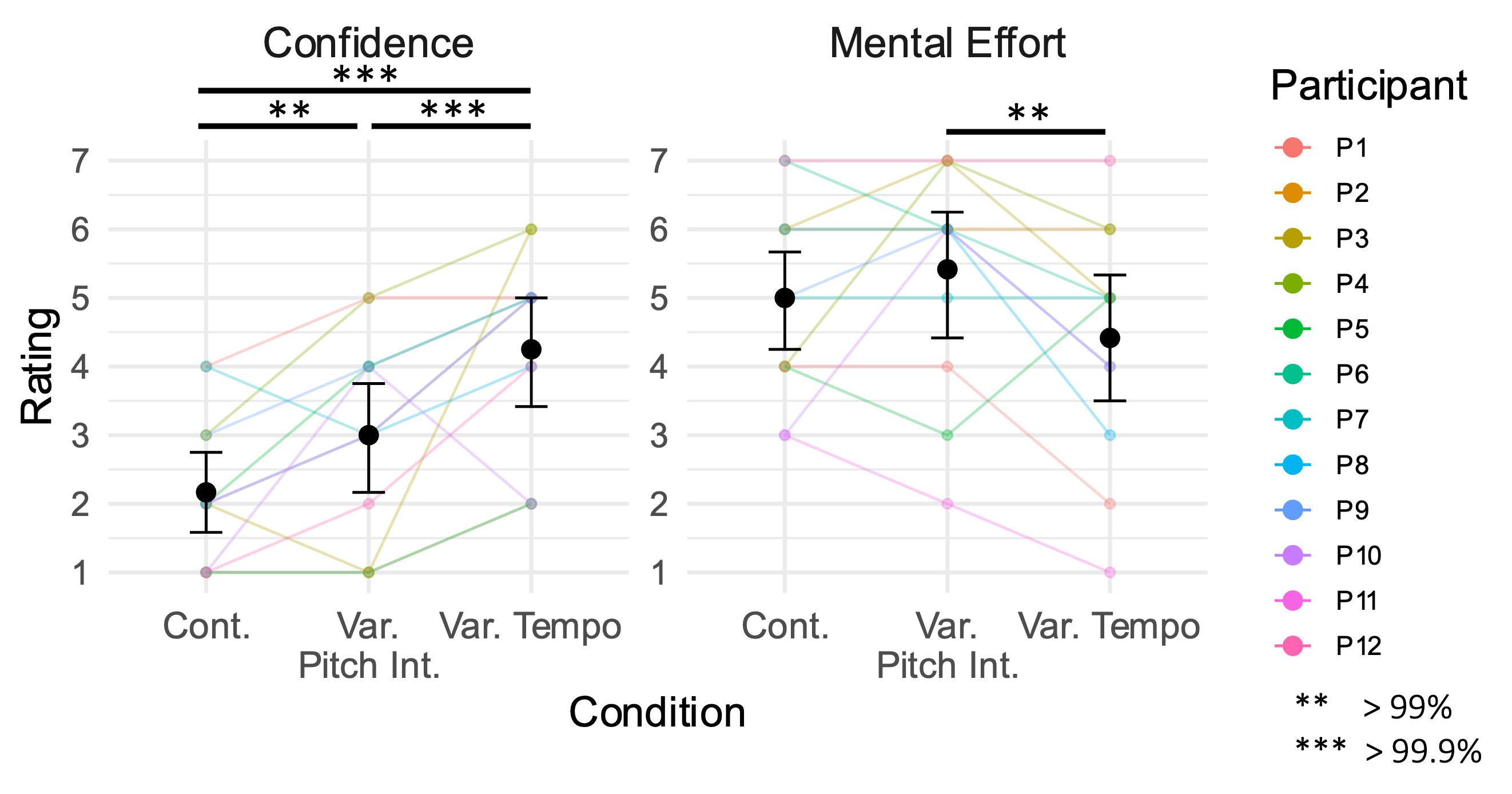}
  \caption{Participant ratings of confidence and mental effort completing slope ratio magnitude estimation task (Experiment 1) across sonification conditions. Higher is better on the left and lower is better on the right. Colored lines represent individual participants (P1–P12), while black points and error bars show group means $\pm 95\%$ bootstrapped confidence intervals. Conditions include Continuous, Variable Pitch Interval, and Variable Tempo. Asterisks above indicate pairwise Bayesian comparisons with posterior probability of direction (PPD): $** > 99\%$, $*** > 99.9\%$.}
  \label{fig:part1Likert}
\end{figure}

\textbf{Confidence:} Participants reported higher confidence under Variable Tempo ($\mu = 4.25, \sigma = 1.48$) than both Variable Pitch Interval ($\mu = 3.00, \sigma = 1.48$) and Continuous ($\mu = 2.17, \sigma = 1.11$) conditions, with differences supported by strong posterior evidence ($PPD > 99.9\%$) (\cref{fig:part1Likert} Left). Relative to the Continuous baseline, the odds of reporting greater confidence were $7.39$ ($95\%$ CI $[1.34, 48.0]$) times higher in the Variable Pitch Interval Condition and $169$ ($95\%$ CI $[17.8, 2170]$) times higher in the Variable Tempo condition.

\textbf{Mental Effort:} Mental effort was rated lowest for Variable Tempo ($\mu = 4.41, \sigma = 1.73$), with greater odds of reporting lower mental effort ($OR = 9.52$, $95\%$ CR $[1.68, 59.2]$, PPD > 99\%) compared to Variable Pitch Interval ($\mu = 5.41, \sigma = 1.62$). (\cref{fig:part1Likert} Right). 

\subsubsection{Strategies and Reflections}

\textbf{Continuous:} Most participants gauged slope magnitude by comparing the start and end of the clip (P3–P6, P9, P10), though fade-ins and fade-outs made this difficult (P4). Others attempted to count note changes (P2, P12), used qualitative labels like “small” or “large” (P2, P7), or relied on intuition (P4, P8, P11). Mental visualization was common---several imagined slope shapes or data plots (P1–P3, P9).

\textbf{Variable Pitch Interval:} Rather than focusing on start and end points, most participants (P4, P5, P8, P10, P12) judged slope by comparing intervals between successive notes. Some used musical terms like “half steps” or “octaves” (P5, P10, P12), while others employed visualization strategies: imagining the data as a plot (P1), visualizing step sizes (P11), or mapping notes to a keyboard (P2). P9 used their palm to sketch or track note changes.

\textbf{Variable Tempo:} All participants relied on tempo, either by judging note speed (P1, P4, P9) or estimating rhythmic subdivisions between clips (P2, P3, P5–P8, P10–P12). Tapping was common (P2, P3, P6, P8, P10–P12), and rhythmic clarity helped---P3 found it easier when patterns subdivided evenly; others (P4, P12) drew on polyrhythmic experience. Only P1 reported using mental visualization, imagining notes on a grid.

\textbf{Most Challenging:} Ten participants (P1–P6, P8, P10–P12) found the Continuous condition hardest due to the lack of discrete reference points and difficulty estimating pitch intervals without segmentation.

\textbf{Most Accurate:} Variable Tempo was reported as most accurate by ten participants (P1–P4, P6–P9, P11, P12), who described tempo cues as easier (P1, P4, P6–P9, P11), more intuitive (P3), and not reliant on visualization (P3). Two participants (P5, P10) favored Variable Pitch Interval for its clarity in pitch judgments.

\textbf{Most Preferred:} The same ten participants (P1–P4, P6–P9, P11, P12) preferred Variable Tempo for its accuracy (P1, P6), low effort (P6, P7, P11), and familiarity (P8, P12).

\subsection{Summary}

Across both quantitative and qualitative analyses, Variable Tempo condition consistently outperformed Variable Pitch Interval and Continuous sampling methods in both accuracy and precision when judging the relative steepness between two slopes. While central tendencies across conditions were relatively similar, underestimation bias was weakest under Variable Tempo as slope ratios grew larger. Additionally, Variable Tempo yielded the lowest residual variability at baseline and was the only condition to maintain stable precision when comparing slopes of opposite signs.

Participants' subjective reports echoed these patterns. Variable Tempo was rated the most accurate and preferred method by the majority of participants, with reliably higher confidence than the other methods and lower mental effort than Variable Pitch Interval. Its use of tempo as a perceptual cue enabled participants to subdivide rhythms or use timing strategies such as tapping. In contrast, Continuous sampling was consistently rated the most difficult due to its reliance on pitch interval judgments and the absence of discrete pitch references.

\section{Experiment 2: Acceleration Discrimination}
\label{sec:exp2}

Detecting acceleration direction (i.e. superlinear vs sublinear growth) is often important in trend interpretation. For sonified functions, this can be quantified as sensitivity to the second derivative of pitch with respect to time. This experiment assessed participants’ ability to detect acceleration under the three sonification sampling methods: Continuous, Variable Pitch Interval, and Variable Tempo. Each auditory stimulus was generated from an accelerating pitch function of the form $pitch(t) = a \cdot t^2 + b \cdot t + c$, where $a$ is acceleration in $semitones/s^2$, $b$ is slope in $semitones/s$, and $c$ is the pitch intercept in $semitones$. We measured the $70.7\%$ just-noticeable difference (JND) threshold \cite{levitt1971transformed} for acceleration (in $semitones/s^2$) and analyzed how it varied with (1) the average slope of the stimulus ($AvgSlope$) and (2) participants’ musical training ($MusicExp$). As in the previous experiment, we collected subjective ratings of confidence and mental effort along with self-reported strategies to contextualize participants’ experiences.

% \begin{table}[h]
%   \caption{%
%     Model variables and their associated hypotheses.%
%   }
%   \label{tab:additional_model_variables}
%   \scriptsize%
%   \centering%
%   \begin{tabu}{%
%     >{\raggedright\arraybackslash}p{0.12\linewidth}%
%     >{\raggedright\arraybackslash}p{0.12\linewidth}%
%     >{\raggedright\arraybackslash}p{0.30\linewidth}%
%     >{\raggedright\arraybackslash}p{0.26\linewidth}%
%   }
%   \toprule
%   \textbf{Variable Name} & \textbf{Parameter Name} & \textbf{Description} & \textbf{Hypothesis} \\
%   \midrule
%   \texttt{AvgSlope} & Average Slope & Average slope in semitones/s. & Curvature would be more difficult to discriminate at high average slope magnitudes. \\
%   \addlinespace
%   \texttt{musicExp} & Years of music training & Self-reported duration of music training. & Sensitivity would improve with music experience based on their dependence on pitch interval judgment. \\
%   \bottomrule
%   \end{tabu}%
% \end{table}

\subsection{Study Procedure and Stimuli}

Participants were first screened via an online survey to ensure no self-reported hearing loss and basic conceptual understanding of acceleration, confirmed by two multiple-choice questions requiring concavity judgments from graphs. The same survey also collected informed consent and demographic data, including age, gender, data and statistical background, sonification experience, and years of musical training. A detailed summary of participant demographics is provided in the Supplemental Material.

Eligible participants completed a 60-minute in-person session, using Audio-Technica ATH-M50x closed-back headphones. As with Experiment 1, the study consisted of a calibration phase, followed by the experimental conditions---Continuous, Variable Pitch Interval, and Variable Tempo---each comprising training, practice, the main task, and a post-task interview. The ordering of conditions was counterbalanced across participants. Participants received a 30USD gift card for participation. All procedures were approved by the Institutional Review Board.

During calibration, the facilitator confirmed audibility by playing the highest (MIDI 88) and lowest tones (MIDI 55) in the study and adjusting volume to the participant’s comfort.

During training, participants received a brief explanation of the mapping and completed a training phase with four example curves varying in slope and concavity. They were asked to identify whether the curves increased or decreased and whether they accelerated upward or downward, with opportunities to ask clarifying questions. 

During practice, participants completed five example trials, each featuring a randomly selected audio clip depicting upward or downward acceleration. After each trial, they indicated the direction of acceleration and received feedback.

The main task used an adaptive staircase with two interleaved 2-down-1-up staircases per condition \cite{levitt1971transformed}---one centered on an average slope of $2$ $semitones/s$, the other on $4$. Each trial presented a single 5-second audio clip with randomly signed acceleration direction (upward or downward) and intercept, and participants indicated the perceived direction. Acceleration began at $1$ $semitone/s^2$ and decreased logarithmically, with step sizes starting at $0.4$ and scaled by $0.7$ after each reversal. Each staircase ended after seven reversals, and the just-noticeable difference (JND) was computed from the last four. Parameters were refined through pilot testing to ensure sensitivity and task balance. To standardize note density across conditions and slope levels, notes were triggered at every semitone change for average slope $4$ and every half-semitone for slope $2$. In the Variable Pitch Interval condition, x-spacings were chosen to match the number of notes produced under Variable Tempo of the same trend.

Each auditory stimulus was rendered with a fixed duration of 5 seconds, consistent with prior psychoacoustic research \cite{Loomis1998Assessing}. After each condition, participants reported their strategy and rated their confidence and mental effort (1–7). At the end, they identified the most challenging, most accurate, and preferred condition, with brief justifications.

All sounds were loudness-normalized using the ISO 226 40-phon contour \cite{suzuki2024}, synthesized with sine oscillators for timbral consistency, constrained to the MIDI 55–88 range for perceptual simplicity \cite{brown2003drawing, wang2022seeing}, and rendered using a custom WebChuck application \cite{chafe2023would}.

\subsection{Participants}

Eighteen sighted individuals (11 men, 7 women), aged 18–34 and not involved in Experiment 1, participated in Experiment 2. Recruited via large university mailing lists, all were prescreened to ensure no self-reported hearing loss and a basic understanding of graphical acceleration.

Only one participant reported (minimal) prior exposure to sonification. Musical training varied: 7 participants had 1–10+ years of formal training, and one reported absolute pitch. 6 had graduate-level and 12 had undergraduate-level statistical coursework.

\subsection{Analysis}

\textbf{Discrimination Performance:} To model acceleration discrimination, we specified a Bayesian hierarchical model (similar to the analysis in Experiment 1) for log acceleration ($log(|a|)$), incorporating features of interest and their interactions with condition \cite{Gelman2013philosophy, Gelman2003bayesian}. We found the log transformation to improve the normality of the data and enable clearer comparisons across conditions.

After exploratory model comparisons, we retained a homoskedastic model that assumes a constant residual variance across all conditions, as incorporating predictors into the sigma formula (e.g., Condition or AvgSlope) did not substantially improve out-of-sample predictive performance based on leave-one-out cross-validation. Effects priors were weakly informative ($\mathcal{N}(0, 5)$). Posterior estimates were consistent across three prior specifications (between $\mathcal{N}(0, 5)$ to $\mathcal{N}(0, 1)$), with maximum absolute differences in posterior means, medians, standard deviations, and interquartile ranges all below 0.0015. The model is specified as follows: $JND_{log(|a|)} \sim 1 + \text{Condition} \times (\text{AvgSlope} + \text{MusicExp}) + (1 | \text{Participant)}$

All Markov chains converged with $\hat{R} \approx 1.00$, and no divergent transitions or sampling pathologies were observed. Posterior predictive checks indicated good model fit, with p-values of $0.808$ (median), $0.508$ (mean), $0.427$ (IQR), and $0.606$ (SD), suggesting the model adequately captured both the central tendency and variability of JND responses. Modeling was conducted using the brms package in R \cite{burkner2017brms}. Models were fit using Hamiltonian Monte Carlo (HMC) with max\_treedepth = 25 and adapt\_delta = 0.98 to improve convergence stability.

\textbf{Participant Ratings:} To evaluate condition effects on confidence and mental effort ratings, we fit Bayesian ordinal regression models using a cumulative logit link and weakly informative priors: $\mathcal{N}(0, 5)$): $\text{Confidence/Mental Effort} \sim \text{Condition} + (1 | \text{Participant})$

\textbf{Strategies and Reflections:} We open-coded participants’ strategies and reflections across conditions, which we summarize in the results.

\subsection{Results}

\subsubsection{Performance}

\begin{figure}[h]% specify a combination of t, b, p, or h for top, bottom, on its own page, or here
  \centering % avoid the use of \begin{center}...\end{center} and use \centering instead (more compact)
  \includegraphics[width=\columnwidth, alt={70.7\% Just Noticeable Difference (JND) estimates for acceleration, measured in semitones per second squared, across three sonification conditions---Continuous, Variable Pitch Interval, and Variable Tempo---at two average slope levels (2 and 4 semitones per second). The y-axis is plotted on a logarithmic scale to reflect perceptual sensitivity, while the x-axis shows the two slope levels. Colored points and connecting lines represent individual participant estimates (P1–P18), highlighting within-subject variation across slope levels. Black dots with error bars indicate the group mean and standard error for each condition and slope level. JNDs generally increased with higher average slope across all conditions, but Variable Tempo consistently resulted in the lowest JNDs, particularly at the lower slope level.}]{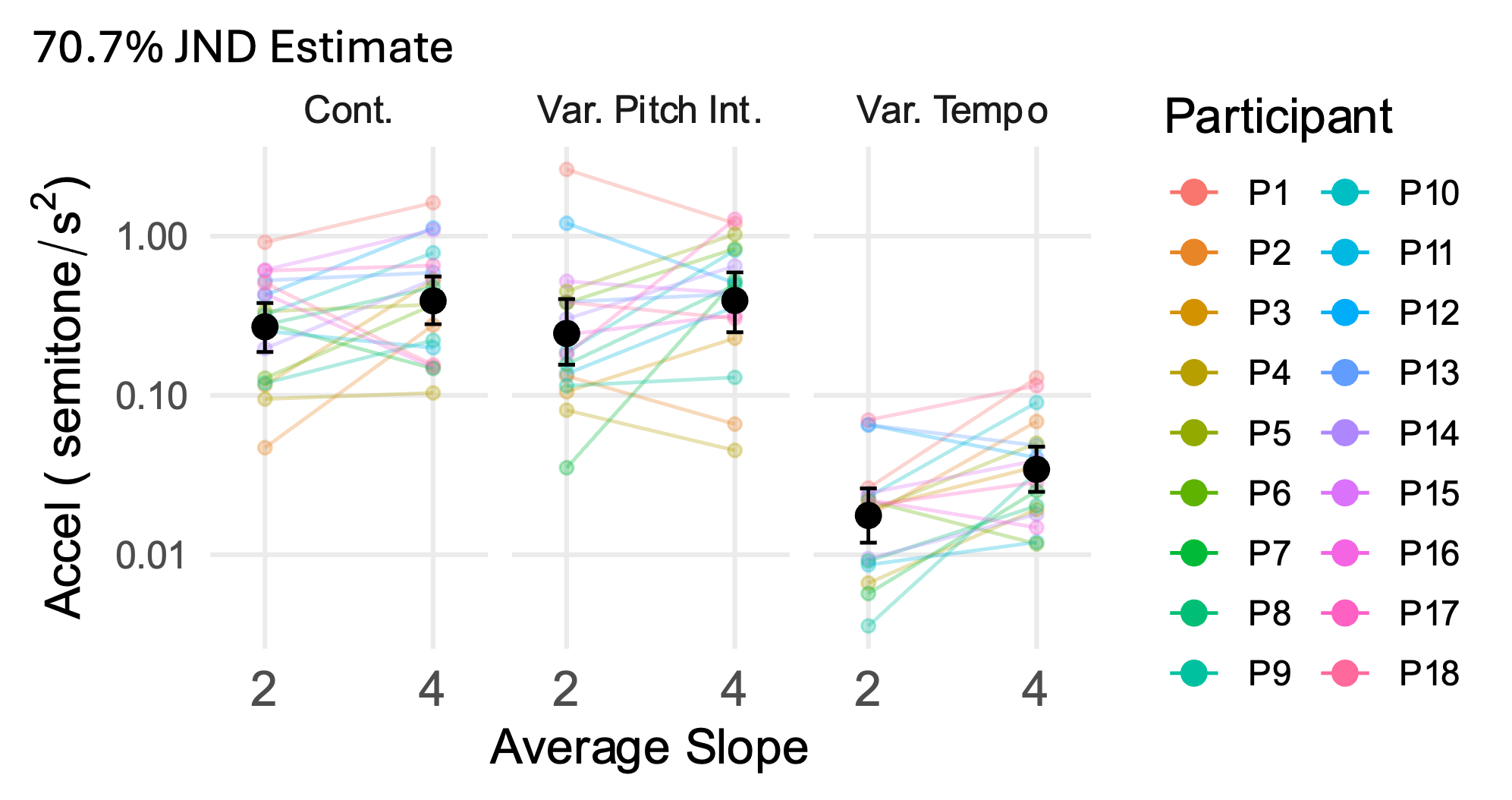}
  \caption{Just-noticeable difference (JND) estimates for acceleration (in $semitones/s^2$) across sonification conditions and two average slope levels ($AvgSlope = 2 \text{ and } 4$). Individual participant thresholds are shown in color, with black points and error bars indicating the group mean and $95\%$ confidence interval. The y-axis is plotted on a logarithmic scale to account for large differences in perceptual thresholds.}
  \label{fig:part2Main}
\end{figure}

\begin{figure}[h]% specify a combination of t, b, p, or h for top, bottom, on its own page, or here
  \centering % avoid the use of \begin{center}...\end{center} and use \centering instead (more compact)
  \includegraphics[width=\columnwidth, alt={Figure titled "Intercept-Adjusted Back-Transformed Estimates for Acceleration JND" showing posterior estimates of just-noticeable difference (JND) thresholds for acceleration discrimination across sonification conditions. Three panels (A to C) display point estimates and 95\% credible intervals on a log-scaled x-axis, with a vertical red dashed reference line at 1.0 for Panels B and C. (A) Baseline JNDs (in semitones/s²) by condition are 0.27 [0.12, 0.51]*** for Continuous, 0.24 [0.11, 0.47]*** for Var. Pitch Int., and 0.01 [0.01, 0.02]*** for Var. Tempo. Triple asterisks indicate >99.9\% posterior probability. Asterisks with bars indicate credible pairwise differences between Variable Tempo and others of >99.9\%. (B) Effect of average slope on JND (as multiplicative factors: 10^β, applied as (10^β)^AvgSlope) for Continuous is 1.21 [0.99, 1.47]*, Var. Pitch Int. is 1.27 [1.04, 1.54]**, and Var. Tempo is 1.40 [1.14, 1.70]***, Asterisks indicate >95\% posterior probability for all conditions. JND increases with slope, most strongly in the Var. Tempo condition. (C) Effect of musical training on JND (as multiplicative factors: 10^β, applied as (10^β)^MusicExp) for Continuous is 0.92 [0.87, 0.98]***, Var. Pitch Int. is 0.90 [0.84, 0.95]***, and Var. Tempo is 0.95 [0.90, 1.01]*. Training is associated with lower JNDs in all conditions, with the strongest effect under pitch-based methods. Vertical bars denote credible differences between Var. Tempo. and Variable Pitch Interval of >95\%}]{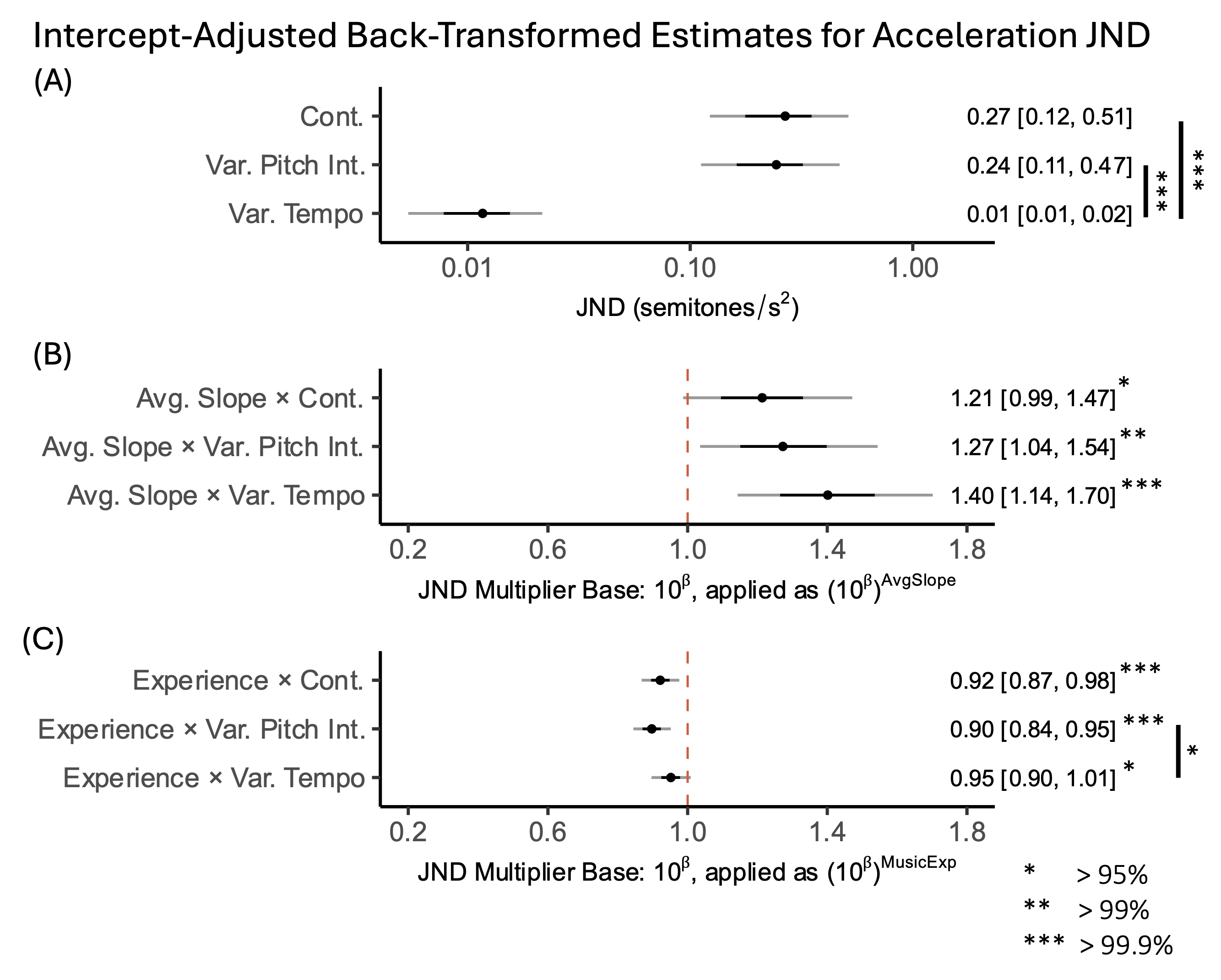}
  \caption{Intercept-adjusted, back-transformed model estimates for acceleration discrimination (JND), along with multiplicative effects of stimulus-level and participant-level features. (A) Estimated just-noticeable difference (JND) in acceleration ($semitones/s^2$) across sonification conditions plotted in log scale. (B) Estimated base multipliers for JND change per unit of average slope ($10^\beta$). Actual modulation is interpreted as $(10^\beta)^{AvgSlope}$. (C) Estimated base multipliers for JND change per unit of musical training ($10^\beta$), interpreted as $(10^\beta)^{MusicExp}$. For base multipliers, values $<1$ indicate increased discrimination sensitivity; values $>1$ indicate decreased discrimination sensitivity. Error bars reflect $66\%$ (gray) and $95\%$ (black) credible intervals. Asterisks denote posterior probability of direction: $* > 95\%$, $** > 99\%$, $*** > 99.9\%$.}
  \label{fig:part2AdjEstimates}
\end{figure}

Across both slope levels, participants could discriminate smaller accelerations in the Variable Tempo ($\mu = 0.0344$, $\sigma = 0.0142$) compared to Variable Pitch Interval ($\mu = 0.480$, $\sigma = 0.638$) and Continuous ($\mu = 0.463$, $\sigma = 0.638$) conditions, reflecting an increase in sensitivity by over 13 times. This pattern was consistent across nearly all individuals (\cref{fig:part2Main}). Posterior estimates confirm this trend, with Variable Tempo yielding the lowest thresholds ($\mu = 0.01 [0.01, 0.02]$), followed by Variable Pitch Interval ($\mu = 0.24 [0.11, 0.47]$) and Continuous ($\mu = 0.27 [0.12, 0.51]$) (\cref{fig:part2AdjEstimates}A).

All conditions showed a credible increase in JND thresholds as average slope increased, meaning that steeper underlying pitch functions made acceleration harder to discriminate. Variable Tempo exhibited the steepest growth, with a base multiplier of $1.40$ ($[1.14, 1.70]$), compared to $1.27$ ($[1.04, 1.54]$) for Variable Pitch Interval and $1.21$ ($[0.99, 1.47]$) for Continuous (\cref{fig:part2AdjEstimates}B).

\begin{figure}[h]% specify a combination of t, b, p, or h for top, bottom, on its own page, or here
  \centering % avoid the use of \begin{center}...\end{center} and use \centering instead (more compact)
  \includegraphics[width=\columnwidth, alt={Figure titled "70.7\% JND Estimate" showing individual and trend-level effects of musical training on acceleration discrimination thresholds (JND) across three sonification conditions: Continuous, Variable Pitch Interval, and Variable Tempo. The x-axis represents years of formal musical training, and the y-axis (log-scaled) represents JND in semitones per second squared. Each panel shows scatter plots of individual participant data with a fitted regression line. In the Continuous condition, a negative slope suggests that more musical training is associated with lower JNDs. In the Variable Pitch Interval condition, this effect is stronger, with a steeper decline indicating greater sensitivity gains with increased training. In the Variable Tempo condition, JNDs are lower overall, and while the trend also declines slightly with training, the slope is shallower, suggesting weaker dependence on musical training.}]{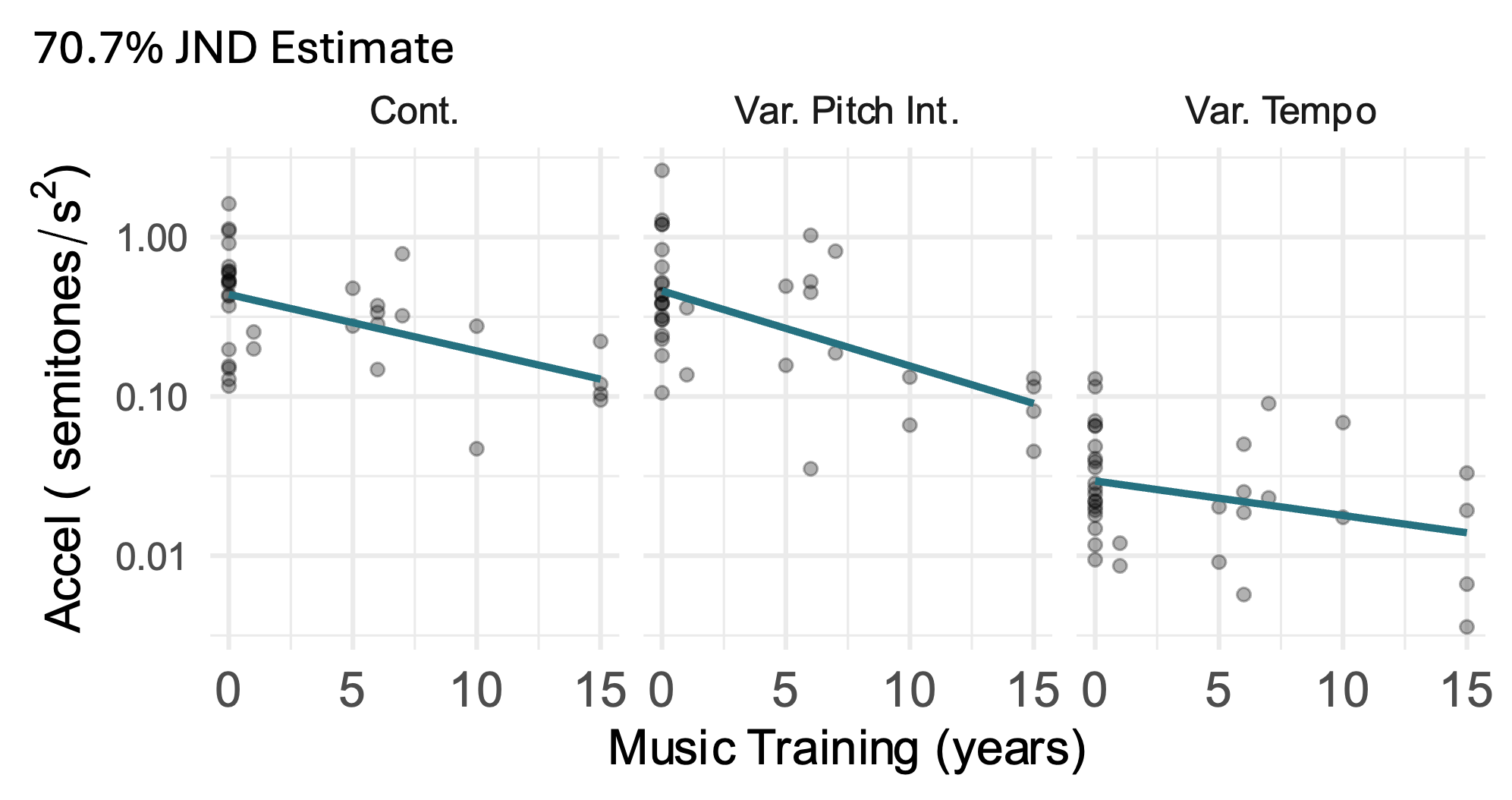}
  \caption{Relationship between musical training and just-noticeable difference (JND) thresholds for acceleration across sonification conditions. Each panel shows JND thresholds as a function of years of formal music training, with regression fits overlaid. The y-axis is plotted on a logarithmic scale to account for large differences in perceptual thresholds.}
  \label{fig:part2Training}
\end{figure}

Musical training was similarly associated with lower JNDs across all conditions (\cref{fig:part2Training}). In each case, $10^\beta < 1$ indicated improved sensitivity with increased training, with the strongest effects in Variable Pitch Interval ($10^\beta = 0.90$ [$0.84$, $0.95$]) and Continuous ($10^\beta = 0.92$ [$0.87$, $0.98$]). The posterior probability that musical training more strongly modulated performance in Variable Pitch Interval than Variable Tempo exceeded $95\%$ (\cref{fig:part2AdjEstimates}C).

\subsubsection{Participant Ratings}

\begin{figure}[h]% specify a combination of t, b, p, or h for top, bottom, on its own page, or here
  \centering % avoid the use of \begin{center}...\end{center} and use \centering instead (more compact)
  \includegraphics[width=\columnwidth, alt={Two line and dot plots showing participant confidence and mental effort ratings across three sonification conditions: Continuous, Variable Pitch Interval, and Variable Tempo. The left plot shows confidence ratings on a 1 to 7 scale. Mean confidence was highest for Variable Tempo and Continuous, and lowest for Variable Pitch Interval. Colored lines trace individual participant responses, while black dots with error bars indicate group means with standard errors. Asterisks above indicate credible differences between Var. Pitch Int. and other conditions with >99\% posterior probability. The right plot shows mental effort ratings on the same 1 to 7 scale. Variable Pitch Interval received the highest average mental effort ratings, followed by Variable Tempo and Continuous. Asterisks above the plot denote credible differences between Variable Pitch Interval and both other conditions with >99.9\% posterior probability. Participant-level ratings are connected by colored lines, with group means marked in black.}]{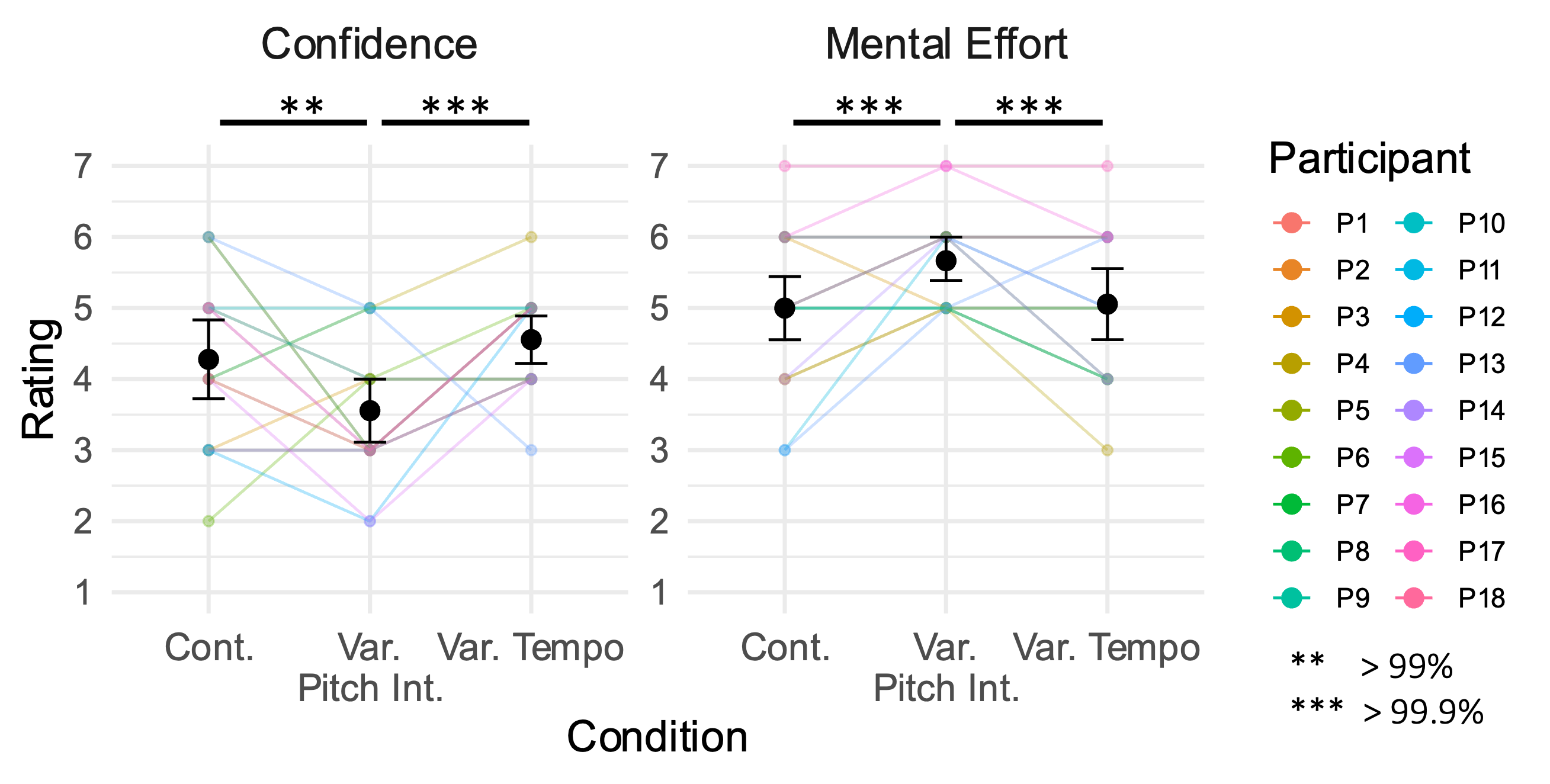}
  \caption{Participant ratings of confidence and mental effort completing acceleration discrimination task (Experiment 2) across sonification conditions. Higher is better on the left and lower is better on the right. Colored lines represent individual participants (P1–P18), while black points and error bars show group means $\pm 95\%$ bootstrapped confidence intervals. Asterisks above indicate pairwise Bayesian comparisons with posterior probability of direction (PPD): $** > 99\%$, $*** > 99.9\%.$}
  \label{fig:part2Likert}
\end{figure}

\textbf{Confidence:} Participants reported lower confidence under Variable Pitch Interval ($\mu = 3.56$, $\sigma = 0.984$) compared to both Variable Tempo ($\mu = 4.56$, $\sigma = 0.705$) and Continuous ($\mu = 4.28, \sigma = 1.18$) conditions. The odds of reporting greater confidence were $7.29$ times higher in Variable Tempo ($95\%$ OR $[1.98, 27.3]$), and $4.31$ ($95\%$ OR $[1.18, 16.5]$) times higher in Continuous relative to Variable Pitch Interval, with strong posterior evidence ($PPD > 99\%$) (\cref{fig:part2Likert} Left).

\textbf{Mental Effort:} Mental effort was also rated highest for Variable Pitch Interval ($\mu = 5.67$, $\sigma = 0.686$), with a notable increase compared to Variable Tempo ($\mu = 5.06$, $\sigma = 1.06$) and Continuous ($\mu = 5.00$, $\sigma = 1.08$) conditions. The odds of reporting greater mental effort were $5.92$ times higher in Variable Pitch Interval compared to Variable Tempo ($95\%$ OR $[1.36, 28.1]$) and $6.36$ ($95\%$ OR $[1.57, 31.2]$) times higher compared to Continuous, with very strong posterior evidence supporting these differences ($PPD > 99.9\%$) (\cref{fig:part2Likert} Right).

\subsubsection{Strategies and Reflections}

\textbf{Continuous:} Most participants gauged acceleration by estimating how long it took for pitch to change (P1, P2, P3, P9, P11–P13, P16), often referencing musical notes or scales (P2, P9), turning the task into one of temporal judgment. Some relied on intuition (P3, P4), visualized slope trajectories (P9, P15), or used metaphors like a ball (P14), car (P15), or siren (P7). One participant (P5) physically embodied the sound by moving their head with the pitch. Attention was split: some compared the beginning and end (P4, P9, P10, P14), others focused only on onset (P6) or ending (P10, P17, P18).

\textbf{Variable Pitch Interval:} Most participants (P1–P7, P9, P11, P15–P16, P18) focused on pitch intervals over time to judge acceleration. P2, who had absolute pitch, found the task more demanding, as their perception tended to “snap all tones to the closest semitone.” Others compared note durations (P8, P12) or visualized steps or balls falling (P10, P12). Attention was again split: some compared the beginning and end (P3–P5, P7, P11, P14), others focused on onset (P6) or the ending (P12, P15–P18), with several noting difficulty recalling the beginning (P12, P17).

\textbf{Variable Tempo:} Nearly all participants (P1–P17) reported judging acceleration based on tempo. Some (P7) kept a rhythm internally while others tapped physically (P6, P15). Similar to the Continuous condition, two participants (P5, P11) judged the relative duration of low versus high-pitched notes. Three participants (P1, P2, P9) also described attending to pitch, though for P9, tempo dominated. Only P18 reported visualizing the curve. P13 used the metaphor of (“traveling on a hill”). Like other conditions, most compared the start and end (P5, P9, P11, P12, P16), while others focused on endings (P6, P10, P13).

\textbf{Most Challenging:} Variable Pitch Interval was most difficult for 10 participants (P5, P7, P8, P10–P12, P14, P15, P17, P18), citing difficulty discerning or tracking pitch changes (P5, P7, P14, P15, P17) or unfamiliarity with pitch (P12). Continuous was hardest for 6 (P1, P3, P4, P6, P9, P16) due to a reported inability to distinguish between pitch intervals (P3, P6, P9), lack of reference (P4), or combined pitch and tempo demands (P16). Two participants (P2, P13) found the Variable Tempo condition most challenging, despite performing best in that condition. P13 noted the trials felt increasingly hard without realizing that trial difficulty adapted to performance, while P2 found tracking changing tempos alongside shifting pitches challenging.

\textbf{Most Accurate:} Variable Tempo was most frequently cited as most accurate (13 participants), described as easier (P3–P5, P12, P14, P15, P17), more intuitive (P9), and compatible with strategies like tapping (P6, P16). Five found Continuous more accurate, citing clarity (P7), certainty (P13), data richness (P2), and compatibility with pitch-based strategies (P1). P10 perceived all methods as equally accurate.

\textbf{Most preferred:} Variable Tempo was ultimately the most preferred (13 participants), praised for ease (P4, P10, P15), intuitiveness (P3, P8, P16), and confidence (P6, P17). Six preferred Continuous (P1, P2, P7, P13, P14, P18), valuing familiarity (P7) and alignment with analogous visual representations (P18).

\subsection{Summary}

Participants demonstrated the greatest perceptual sensitivity in the Variable Tempo condition, requiring the smallest acceleration differences to detect change. Across all conditions, thresholds increased with higher average slope, with the steepest rise observed for Variable Tempo. Musical training was associated with lower thresholds overall, with the strongest effect seen in the Variable Pitch Interval condition.

Subjective feedback was mostly consistent with performance results. Variable Tempo was most frequently rated as the easiest, most accurate, and most preferred condition, attributed to the availability of timing cues that aligned with participants’ perceptual strategies. Surprisingly, despite yielding credibly higher discrimination thresholds, continuous sampling was also favored by many, receiving comparable ratings for confidence and mental effort. In contrast, Variable Pitch Interval was often described as more mentally taxing and harder to follow, a view reflected in elevated effort ratings and supported by strong posterior evidence.

\section{Discussion}

Among the three methods tested, Variable Tempo sampling yielded lower estimation error, finer acceleration JNDs, and more favorable ratings. Here, we situate these findings within perceptual and visualization research and outline directions for future work.

\subsection{Pitch and Tempo Perception}

\subsubsection{Slope and Acceleration Judgment Using Pitch Intervals}

Our findings align with and extend established psychophysical research on pitch and tempo perception. Lower performance in Variable Pitch Interval and Continuous conditions reflects longstanding evidence that, although humans have fine frequency resolution, pitch interval discrimination is imprecise and cognitively demanding \cite{mcdermott2010musical, little2019inducing, burns1978categorical, foxton2004training}. 
% Pitch-based sonification has also been shown to introduce substantial error in magnitude estimation, despite outperforming other audio channels like length or volume \cite{wang2022seeing}.

Slope ratio magnitude estimation in Experiment 1 introduced several layers of added complexity beyond single-interval discrimination. First, interval judgments under Continuous sampling---described by many participants as the most difficult---lacked discrete reference points (i.e., note onsets). Second, participants had to compare two moving sequences of pitch intervals. Larger slope ratios were increasingly underestimated, consistent with findings on the compression of large pitch intervals \cite{russo2005subjective} and numerical magnitudes \cite{dehaene2008log, fechner1860elemente}.

We found limited benefits of musical training for slope ratio magnitude estimation accuracy, aligning with prior findings in sonification research \cite{walker2011theory, watson1994factors}. One potential factor may be that the working memory demands of comparing evolving pitch structures across two sequences may overshadow the utility of prior training.

In Experiment 2, which required detecting acceleration through the temporal expansion or contraction of pitch intervals, pitch-based conditions again resulted in lower sensitivity. Performance followed a Weber-like pattern, with thresholds increasing alongside average slope (i.e., interval size). Here, musical training was more beneficial, aligning with prior findings of musical training improving pitch interval discrimination \cite{mcdermott2010musical}.
% suggesting that second-order auditory judgments involving dynamic pitch changes may be more sensitive to formal experience.

Interestingly, while the Continuous condition was most often rated as the most difficult in Experiment 1 due to the absence of discrete reference points, Pitch Interval was most often rated as the most difficult in Experiment 2. Many cited difficulty remembering and comparing pitch interval sequences, echoing Deutsch’s findings on interference from intermediate tones \cite{deutsch1970tones}. Some reported adapting Continuous sampling into a timing-based strategy that aided performance. Taken together, these findings suggest that pitch may be suboptimal for tasks requiring derivative features like slope or acceleration, particularly considering perceptual or memory limits \cite{mcdermott2010musical, deutsch1970tones}.

\subsubsection{Slope and Acceleration Judgment Using Tempo}

While pitch-based strategies presented perceptual challenges, participants consistently performed best under the Variable Tempo condition. One possible explanation is the use of intervals aligned with Western musical scales in the study. However, prior studies have not found reliable perceptual advantages for intervals aligned with Western tuning over those that are not \cite{mcdermott2010musical}. Moreover, musical experience had minimal impact on performance in Experiment 1, and in Experiment 2, participants performed better with half-semitone than full-semitone steps. 
% Taken together alignment with Western tuning is unlikely to be the primary factor driving the observed effects.

Another explanation for improved performance is the incorporation of temporal cues, supported by both our findings and prior perceptual research. Participants frequently emphasized relying on timing as a primary strategy. Unlike pitch interval comparisons, which can be imprecise and memory-dependent, tempo judgments are relatively accurate even among untrained listeners and particularly precise among those with rhythmic expertise, such as percussionists or DJs \cite{foster2021accuracy}. Temporal perception also benefits from redundancy through the multiple-look effect, which may enhance discrimination \cite{drake1993tempo}.

In Experiment 1, slope ratio estimation involved comparing the relative tempo of two note sequences. High performance may reflect the auditory system’s sensitivity to metrical structure, enabling grouping into simple integer ratios such as 2:1 \cite{moller2021beat}. Participants commonly reported entraining to one sequence and mentally overlaying or subdividing the other, relying on metrical relationships rather than pitch cues. These strategies appeared broadly accessible regardless of musical background, though slight underestimation of extreme ratios persisted.

In Experiment 2, acceleration discrimination required detecting drift from a steady tempo. Participants showed over 13-fold finer just-noticeable differences (JNDs) under Variable Tempo compared to pitch-based methods, highlighting the auditory system’s sensitivity to gradual tempo changes \cite{madison2004detection, dahl2003estimating}. Performance declined with faster baseline tempi, consistent with Weber-like scaling \cite{thomas2007just}, and improved with musical training---suggesting a learned component in tempo sensitivity.

A likely contributor to the observed two-fold JND increase between average slopes of 2 and 4 is the inter-onset interval adjustment: to maintain a constant number of notes, we doubled the pitch step size, halving both baseline tempo and the tempo rate of change as average slope increased (see \cref{sec:VariableTempo}). This produced a flatter temporal gradient in AvgSlope = 4 trials that likely reduced cue salience. Future work could investigate whether shorter inter-onset intervals reliably enhance discrimination or approach perceptual limits at extreme tempi.

By aligning sonification design with rhythmic abilities, Variable Tempo may offer a perceptually grounded and user-preferred sampling method. Nonetheless, further research is needed to disentangle the individual contributions of temporal cues and scale structure. Additionally, exploring how pitch–tempo interactions \cite{pazdera2025pitch, duke1988effect} influence sonification judgments represents an important direction for future work.

\subsection{Implications for Practice}

\subsubsection{Applying Sampling Methods in Practice}

The experimental conditions in this study assume an underlying continuous function that can be discretized using either uniform x- or y-spacing, enabling the Variable Pitch Interval and Variable Tempo methods, respectively. While this assumption holds for many mathematical contexts (e.g., standard functions, probability distributions), real-world data are often pre-sampled or aggregated. Extending Variable Tempo Sampling to discontinuous data requires additional processing, as described below.

\textbf{Data collection:} If the underlying phenomenon is continuous (e.g., temperature, acceleration), one approach is to collect data at both fixed time intervals and predefined y-axis changes. This parallels event-triggered strategies used in ECG or Geiger counters, where time intervals vary in response to signal change.

\textbf{Data processing:} For non-continuous functions, Variable Tempo can be implemented by fitting a continuous curve (e.g., spline) and resampling it at uniform y-intervals to generate tempo-varying note timings. If preserving original data points is critical, additional notes can be interpolated between them based on y-interval crossings. These interpolated notes may be softened or timbrally differentiated to distinguish them from true data points. To ensure perceptual clarity---particularly in steep regions---y-step size or playback tempo may need adjustment to avoid excessive note density.

Importantly, all three sampling methods convey the same overall pitch trajectory, enabling strategy selection to be guided by goals such as precision, interpretability, or personal preference.

\subsubsection{Visualization and Multimodal Data Use}

Our findings introduce several potential opportunities in the context of visualization, multimodal analysis, and auditory display design.

Experiment 1 showed that Variable Tempo Sampling can support higher precision and accuracy in participants’ estimates of slope ratios compared to other pitch-based mappings. This may be useful not only for examining changes within a single function (e.g., time series data) but also for comparing across functions, such as understanding how time constants affect exponential decay.

These temporal cues may also complement visual graphs. Since angle-based slope estimation is prone to error, especially at steeper slopes \cite{talbot2012empirical}, encoding rates through timing rather than geometry may yield more consistent judgments across a broader range of values.

Experiment 2 demonstrated that participants can detect subtle differences in acceleration under the Variable Tempo condition. Such sensitivity could be relevant in tasks where distinguishing between linear and nonlinear trends is important, such as function learning, motion analysis, or population modeling.

In real-time monitoring, systems like ECGs use tempo to convey rate changes in one-dimensional data. Our work extends this to two-dimensional sonification by varying both pitch and tempo to represent magnitude and rate of change between variables. This approach may support awareness of direction and rate of change when visual attention is limited, such as signaling temperature at each degree shift of a power plant or altitude at each meter change of a flight system.

Because tasks in this study were intentionally constrained to isolate perceptual mechanisms, further research is needed to assess the utility and interpretability of these mappings across these contexts.

\section{Limitations and Future Work}

% While this study provides novel insights into how different sonification sampling strategies affect perceptual judgments of slope and acceleration, 

Several limitations suggest important directions for future research.

\textbf{Participant Representation:} The participant pool consisted primarily of sighted college-aged individuals with backgrounds in data analysis or related fields. While this group reflects a relevant demographic for data-intensive contexts, there is a need to evaluate performance across more diverse populations, particularly those for whom auditory displays may offer the greatest accessibility benefit. In particular, investigating how these findings generalize to blind and low-vision users, accounting for variation in education, data literacy, and onset of visual impairment, represents an important direction for future work.
% For example, while blind and low-vision users exhibit similar scaling functions and polarity preferences when relating data values to acoustic features, data literacy and prior exposure to visualization concepts may differ between individuals who are congenitally blind and those who lost vision later in life.

\textbf{Stimulus Design:} All auditory stimuli were limited to five-second, one-shot audio clips. While this design enabled controlled comparisons across conditions, results may differ in real-world scenarios where data excerpts vary in length and users can replay, scrub, or interactively explore the data \cite{zhang2024charta11y, chundury2023tactualplot}. Additionally, while Madison et al. \cite{madison2004detection} did not find any credible differences in discrimination threshold between increasing and decreasing intervals, Experiment 2 only examined functions with positive overall slopes. Investigating performance across multiple listens, active exploration (e.g., touch-based navigation or scrubbing), and multiple slopes can inform a broader range of real-world scenarios.

\textbf{Sound Design and Timbre:} All study cues used sine oscillators for clarity and minimal spectral complexity. However, certain sonifications often involve more harmonically rich timbres. Extending this work to different oscillators or complex sounds to examine effects on discriminability or listener preference is another area for future work.

\textbf{Task Scope:} We focused on two foundational perceptual tasks: estimating slope ratio magnitudes and discriminating acceleration. While important, they do not capture the full range of insight-seeking behaviors involved in data analysis. Future research examining how sampling strategies impact other types of analytical tasks could provide a more comprehensive understanding of effective sonification design.

% \textbf{Modeling Approach:} This study used Bayesian hierarchical models to estimate both mean accuracy and residual variability across conditions and covariates. While this approach allowed for principled comparison and uncertainty quantification, the sample size constrained our ability to explore higher-order interactions. Larger participant pools could support more granular modeling of covariates such as musical training, cognitive strategy, or trial-level uncertainty.

% \textbf{Cognitive and Affective Factors:} Finally, while we measured subjective mental effort and confidence, additional research could examine cognitive load, learning effects, and emotional response to different sonification strategies over time. Understanding how listeners adapt to or internalize sonification mappings could inform training protocols and the design of more intuitive sonification systems.

\section{Conclusion}

This study investigated how different sonification sampling strategies influence users’ ability to perceive slope and acceleration from pitch-based sonifications of functions. Across two experiments, Variable Tempo sampling---a novel method in which tempo emerges naturally from uniform sampling along the y-axis---consistently outperformed traditional approaches. When comparing slopes, participants using Variable Tempo made less biased estimates at higher slope ratios and exhibited greater precision, especially when judging slopes with opposite signs. They were also able to detect significantly smaller differences in acceleration under this condition. Participants frequently relied on rhythmic strategies such as tapping and temporal grouping, suggesting that emergent temporal patterns may provide perceptual advantages beyond pitch variation alone. These findings highlight sampling-driven temporal structure as a promising design mechanism for improving perception and task performance related to derivative features of functions.

%% if specified like this the section will be omitted in review mode
\acknowledgments{This research was supported by the Stanford Institute for Human-Centered AI, NSF Award 2016789, and by an unrestricted gift from Reality Labs Research, a division of Meta.}

\bibliographystyle{abbrv-doi-hyperref}

\bibliography{template}

\appendix % You can use the `hideappendix` class option to skip everything after \appendix

\end{document}